\documentclass[fleqn,usenatbib]{mnras}

\usepackage{newtxtext,newtxmath}

\usepackage[T1]{fontenc}

\DeclareRobustCommand{\VAN}[3]{#2}
\let\VANthebibliography\thebibliography
\def\thebibliography{\DeclareRobustCommand{\VAN}[3]{##3}\VANthebibliography}

\usepackage{graphicx}	
\usepackage{amsmath}	
\usepackage{mathtools}
\usepackage{lineno}
\usepackage{lscape}

\title[Two-Screen Scattering in CRAFT FRBs]{Two-Screen Scattering in CRAFT FRBs}

\author[M.~W.~Sammons et al.]{
Mawson~W.~Sammons,$^{1}$\thanks{E-mail: mawson.sammons@postgrad.curtin.edu.au}
Adam~T.~Deller,$^{2}$
Marcin~Glowacki,$^{1}$
Kelly~Gourdji,$^{2}$
C.~W.~James,$^{1}$
\newauthor
J.~Xavier~Prochaska,$^{3,4,5}$
Hao~Qiu,$^{6}$ 
Danica~R.~Scott,$^{1}$
R.~M.~Shannon,$^{2}$
and C.~M.~Trott,$^{1}$
\\
$^{1}$International Centre for Radio Astronomy Research, Curtin University, Bentley, WA 6102, Australia\\
$^{2}$Centre for Astrophysics and Supercomputing, Swinburne University of Technology, Hawthorn, VIC, 3122 Australia\\
$^{3}$Department of Astronomy and Astrophysics, University of California, Santa Cruz, CA 95064, USA\\
$^{4}$Kavli Institute for the Physics and Mathematics of the Universe, 5-1-5 Kashiwanoha, Kashiwa 277-8583, Japan\\
$^{5}$Division of Science, National Astronomical Observatory of Japan, 2-21-1 Osawa, Mitaka, Tokyo 181-8588, Japan\\
$^{6}$SKA Observatory, Jodrell Bank, Macclesfield, SK11 9FT United Kingdom\\
}

\date{Accepted XXX. Received YYY; in original form ZZZ}

\pubyear{2021}

\begin{document}
\label{firstpage}
\pagerange{\pageref{firstpage}--\pageref{lastpage}}
\maketitle

\begin{abstract}
Temporal broadening is a commonly observed property of fast radio bursts (FRBs), associated with turbulent media which cause radiowave scattering. Similarly to dispersion, scattering is an important probe of the media along the line of sight to an FRB source, such as the circum-burst or circum-galactic mediums (CGM). Measurements of characteristic scattering times alone are insufficient to constrain the position of the dominant scattering media along the line of sight. However, where more than one scattering screen exists, Galactic scintillation can be leveraged to form strong constraints. We quantify the scattering and scintillation in 10 FRBs with 1) known host galaxies and redshifts and 2) captured voltage data enabling high-time resolution analysis. We find strong evidence for two screens in three cases. For FRBs 20190608B and 20210320C, we find evidence for scattering screens less than approximately 16.7 and 3000\,kpc respectively, from their sources, consistent with the scattering occurring in the circum-burst environment, the host ISM (inter-stellar medium) or the CGM. For FRB\,20201124A we find a low modulation index that evolves over the burst's scattering tail, indicating the presence of a scattering screen $\approx9$\,kpc from the host, and excluding the circum-burst environment from potential scattering sites. By assuming that pulse broadening is contributed by the host galaxy ISM or circum-burst environment, the lack of observed scintillation in four FRBs in our sample suggests that existing models may be poor estimators of scattering times associated with the Milky Way's ISM, similar to the anomalously low scattering observed for FRB\,20201124A.
\end{abstract}


\section{Introduction}\label{sec:intro}
Fast radio bursts (FRBs) are short duration ($\mu$s -- ms), extragalactic, radio frequency bursts \citep{lorimer_bright_2007, thornton_population_2013}. In addition to intrinsic time-frequency structure, FRBs are dispersed and often contain the hallmarks of multi-path propagation, arising from propagation through a turbulent medium. While the intergalactic medium (IGM) is often responsible for a sizeable portion of FRB dispersion \citep{macquart_census_2020}, due to its tenuous density, it is not expected to contribute significantly to the scattering, with estimates typically as low as $\sim10\,\mu$s at $1$\,GHz \citep{macquart_temporal_2013, cordes_redshift_2022}. This conclusion is supported by the observed lack of correlation between FRB dispersion measures (DM) and scattering times \citep{chawla_modeling_2022, gupta_ultranarrow_2022}.

Similarly, the Milky Way interstellar medium (ISM) is not expected to dominate the scattering observed in FRBs at high Galactic latitudes, with scattering times inferred from pulsars \citep{cordes_ne2001i_2003} being $\lesssim10\mu s$ for lines of sight more than $30^\circ$ away from the Galactic plane. Assuming that the host galaxies of FRBs are similar to the Milky Way, the symmetry of the scattering process leads to the conclusion that, on average, host galaxy ISMs are also unlikely to be singularly responsible for the observed FRB scattering \citep{simha_disentangling_2020, chawla_modeling_2022}. 

Due to their large geometric leverage, intervening galaxies are a potential source of large scattering in FRBs. For a high redshift population ($z\sim5$), intervening galaxies have been forecast to be the dominant source of scattering \citep{ocker_scattering_2022}. For FRBs with $z\lesssim1$, however, the probability of intersecting a foreground galaxy is insufficient for them to be the dominant source of scattering in the population \citep{macquart_temporal_2013, prochaska_astrophysical_2018, chawla_modeling_2022, ocker_scattering_2022}. 

A potentially important scattering region is within the circum-burst environment, which has long been suggested as the site of the $\gtrsim$\,ms scattering times and $\gtrsim\,100$\,rad\,m$^{2}$ rotation measures (RMs) observed in some FRBs \citep{masui_dense_2015}. Measurements of RM variability in some repeating FRBs have supported this scenario, with large variations over short durations requiring a dense, magnetised medium near the source \citep{michilli_extreme_2018, hilmarsson_rotation_2021, anna-thomas_highly_2022}. Recent measurements of scattering variability in FRB\,20190520B provide the tightest limits yet, with variation on minute timescales requiring the dominant scattering media to be within at most 0.4\,AU of the source, and potentially within $\sim10^4$\,km \citep{ocker_scattering_2022}. In this scenario, scattering serves as an important probe of the circum-burst region which would inform our understanding of FRB progenitors, favouring formation channels where the central engine evolves in a dense turbulent magnetised medium, such as a magnetar embedded within a nebula \citep{margalit_concordance_2018}.

Another region of interest is the circum-galactic media (CGM) of foreground galaxies. To date, observations of FRBs passing through the CGM/halos of intervening galaxies have shown very little scattering, with only as much as $\sim80\,\mu$s recorded at $1.4$\,GHz \citep{prochaska_low_2019, connor_bright_2020, connor_deep_2023}. The possible presence, however, of cloudlets of cold gas in the CGM, inferred from quasar absorption spectra \citep{mccourt_characteristic_2018}, has the potential to cause scattering consistent with that observed in the FRB population \citep[][see \cite{prochaska_low_2019} for a corrected description]{vedantham_radio_2019}. If this model is correct, then FRBs could serve as an important probe of the CGM. As discussed by \cite{vedantham_radio_2019}, distinguishing between scattering in the CGM and circum-burst media will be crucial. 

For scattering that is well approximated by a thin screen model, the degeneracy between the angular broadening and screen distance makes it difficult to directly constrain where the scattering is occurring based only on the pulse-broadening time. For repeating FRBs, a direct constraint can be made by observing the variation of decorrelation bandwidth ($\nu_{\text{DC}}$) or temporal broadening over time ($t_{\text{scatt}}$) \citep{ocker_scattering_2022, main_modelling_2022}. For FRBs that are not seen to repeat, a variability study cannot be conducted; however, in cases where scattering and scintillation have been contributed by separate screens, the scattering geometry can be constrained using the observation of only a single burst \citep{masui_dense_2015, farah_frb_2018, ocker_large_2022}. This can allow not only for the distinction between host and intervening scattering screens, but also constrain the level of scatter-broadening in the Milky Way, for independent comparison with electron distribution models such as NE2001 and YMW16 \citep[respectively]{cordes_ne2001i_2003, yang_dispersion_2017}. 

Where previously \cite{day_high_2020} relied on lower time resolutions and image-plane-based techniques, it is now routinely possible to conduct detailed burst morphology analysis, of the type undertaken by \cite{cho_spectropolarimetric_2020}, for all Commensal Real-time ASKAP (Australian Square Kilometre Array Pathfinder) Fast Transient survey (CRAFT) FRBs with the advent of the CELEBI post-processing pipeline \citep{scott_celebi_2023}. This allows for the high-precision estimates of $\nu_{\text{DC}}$ and $t_{\text{scatt}}$, required to robustly identify scintillation and scattering. We are therefore motivated to search for evidence of two-screen scattering within CRAFT FRBs. 

In this work we measure the level of scattering and scintillation in 10 CRAFT FRBs with high spectro-temporal resolution and apply the two-screen model developed by \citet{masui_dense_2015} and \citet{ocker_large_2022} to place constraints on the distances to their respective scattering screens. In \S \ref{sec:method} we detail the data and our methodology. In \S \ref{sec:results} we present the results, and in \S \ref{sec:discussion} we discuss their implications.

\section{Method}\label{sec:method}             

The scattering and scintillation resulting from multi-path propagation through the same medium will be related via a Fourier uncertainty relationship,
\begin{equation}\label{eq:uncertainty}
    2\pi\nu_{\text{DC}}t_{\text{scatt}}=C.
\end{equation}
The precise value of $C$ depends on the geometry and the density fluctuations in the scattering media, however, it typically ranges between 0.5 and 2 \citep{lambert_theory_1999}. As observed previously by \citet{masui_dense_2015} and \citet{ocker_large_2022}, this is not always the case for FRBs, with discrepancies indicating that a single scattering medium is a poor model for propagation along the line of sight. 

\begin{figure}
    \centering
    \includegraphics[width=0.5\textwidth]{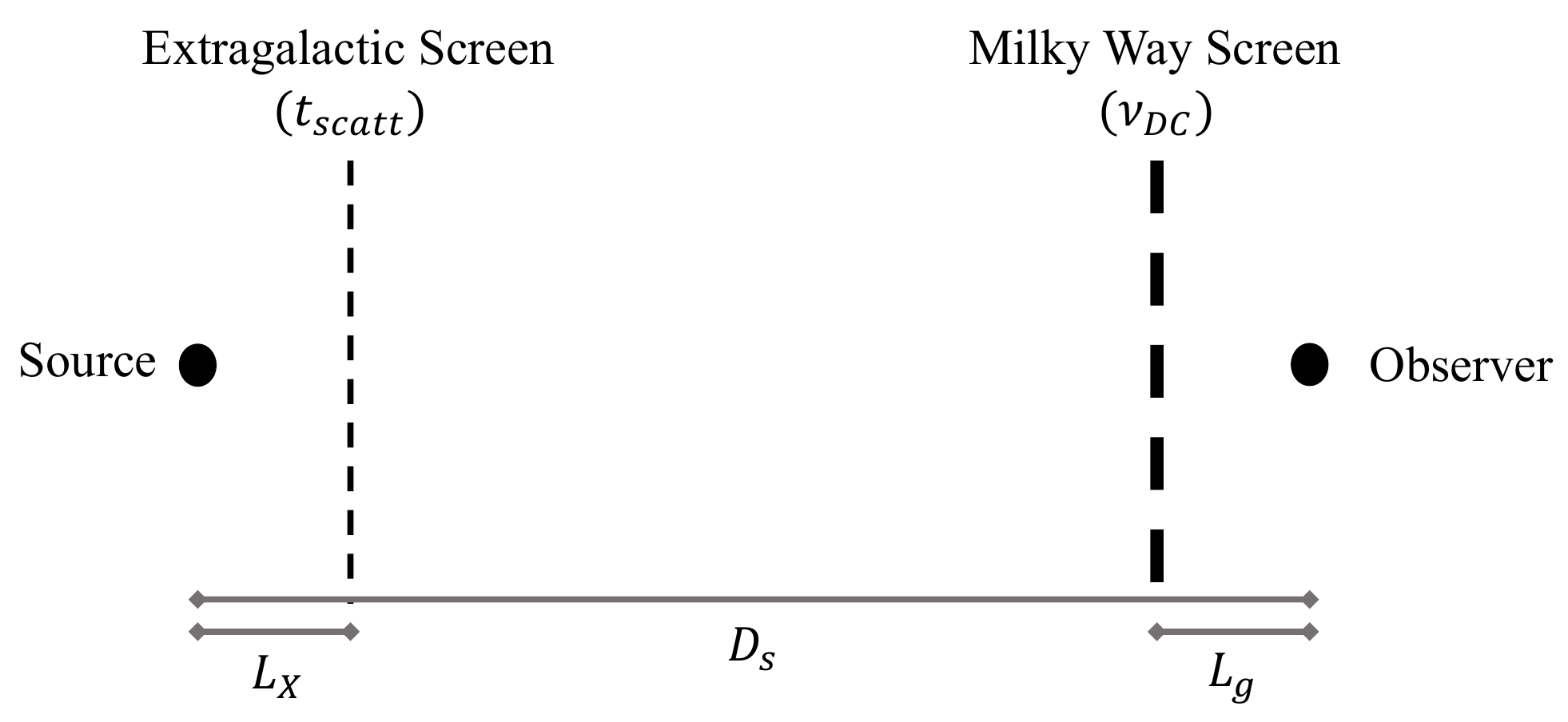}
    \caption{Diagram of the two screen scattering geometry.}
    \label{fig:geomDiagram}
\end{figure}

In these cases, a two-screen model can provide a natural explanation for the differences. Under this model, a relatively large $t_{\text{scatt}}$ and $\nu_{\text{DC}}$ are contributed by separate screens, allowing them to be observed simultaneously for a given line of sight without violating the uncertainty relationship within a single screen. The geometry is often described as shown in Fig. \ref{fig:geomDiagram}, using $L_x$, the distance between the source and the first screen, and $L_g$, the distance between the observer and the second screen, labeled according to the expectation that the first screen is extragalactic and the second Galactic. In order for both screens to cause diffractive scintillation, the scattered image formed by the first screen must be unresolved by the second. Assuming that the distance between the scattering screens is much larger than either $L_x$ or $L_g$ (i.e. $(L_x+L_g)/D_s\ll1$), this leads to the following constraint on the geometry \citep{ocker_large_2022}
\begin{equation}\label{eq:twoScreenDist}
     L_xL_g\lesssim \frac{D_s^2}{2\pi\nu^2(1+z)}\frac{\nu_{\text{DC}}}{t_{\text{scatt}}}.
\end{equation}
The factor of $(1+z)$, where $z$ is the redshift of the source, results from using the cosmological relation between $t_{\text{scatt}}$ and $\theta_{\text{scatt}}$ as derived in \cite{macquart_temporal_2013}. This allows the position of scattering media to be constrained directly using once-off FRBs, allowing mediums such as a diffuse IGM to be ruled out as the dominant source of scattering. In the case of FRB\,201905020b, a reasonable assumption of $L_g$ places the extragalactic screen within $100$\,pc of the FRB progenitor, suggesting that the scattering could be occurring in the circum-burst environment \citep{ocker_large_2022}. 

Our data comprise 10 localised FRBs which had been processed through the CELEBI post-processing pipeline \citep{scott_celebi_2023} at the time of writing. These bursts were detected in real-time searches of the incoherent sum of intensities of each antenna in each of the 36 beams formed digitally using ASKAP's phased-array receivers \citep{bannister_detection_2017, bannister_single_2019}. Each detection triggered the download of the $3.1$\,s voltage buffers channelised using an oversampled polyphase filterbank (PFB). To localise the FRB, the voltage data are correlated, calibrated and imaged as detailed by \citet{day_astrometric_2021} and exemplified in \citet{ryder_probing_2022}. To study the burst morphology, as we shall here, the PFB is inverted to recover the full $\sim3\,$ns time resolution of the voltage data, which are then beamformed and dedispersed as outlined in \citet{scott_celebi_2023}, with the dispersion measure (DM) chosen to optimise the sharpness of temporal structures within the bursts as detailed in \citet{sutinjo_calculation_2023}. We identify the bursts within the $\sim3$\,s of voltage data sampled at $\sim3$\,ns resolution and form dynamic spectra, using a Fast Fourier Transform, of the four Stokes parameters. By default the resolution is chosen to be $0.1$\,MHz and $10\,\mu$s. However, when temporal or spectral structures were found to be unresolved, respective scales as small as $1\,\mu$s or $10$\,kHz were explored independently\footnote{In these cases where greater resolution was required the spectral and temporal analyses were performed on separate data sets formed from the same voltages at independent resolutions}. Here we analyse only the Stokes I data associated with each burst; a more complete polarimetric study of each burst is reserved for a future work.

From the dynamic spectra formed, we select on and off-pulse (pre-burst) regions to account for the shape of the bandpass and to mitigate any radio frequency interference (RFI). To do this, the time-averaged spectrum in the off-pulse region is subtracted from the burst and each spectral channel in the burst is divided by the standard deviation of the corresponding off-pulse channel. The resulting burst dynamic spectrum has a noise that is normally distributed with a mean of zero and a standard deviation of unity. Furthermore, the burst intensity is now represented in units of per-channel signal-to-noise ratio (S/N). 

The level of spectral modulation in a burst is calculated from the lowest, non-zero frequency lag in the mean normalised spectral auto-covariance function (ACF), as per \cite{macquart_spectral_2019}. Bursts with a high modulation index (m) or obvious scintillation in their dynamic spectra are then investigated further. Following other studies \citep{nimmo_burst_2022, ocker_large_2022}, we fit a Lorentzian to the ACF of the mean-subtracted, time-integrated, normalised burst spectra and we measure the $\nu_{\text{DC}}$ to be the half-width-half-maximum (HWHM) of the best-fit case. 

In cases where significant RFI is present in the unnormalised burst dynamic spectrum, we investigate the impact of RFI subtraction on the ACF of normalised bursts. To do so, a fake FRB with a uniform spectral profile is injected into the off-pulse noise and then normalised via the same method. If a significant excess is found in the ACF of this normalised fake FRB then the RFI is deemed too significant to compensate for, and the FRB in question (or at least the section of bandwidth containing the RFI) are discarded from the sample. To avoid large Poisson noise associated with measuring only a small number of scintles (the finite scintle effect) \citep{cordes_timing_1990}, we require the retained bandwidth to be much larger than $\nu_{\text{DC}}$.

To distinguish scintillation from frequency structures intrinsic to the burst such as self-noise\footnote{Following \citet{ocker_scattering_2022}, we refer to frequency structures on the reciprocal scale of FRB temporal sub-structures as self-noise.}, we split the normalised FRB into four even sub-bands and fit a Lorentzian to the ACF of each band's spectrum. Due to the large number of scintles in each sub-band we expect the effect of re-binning on the results will be minimal. If $\nu_{\text{DC}}$ is observed to increase with frequency, as expected for multi-path propagation through a cold plasma, we assume the spectral structures are caused by scintillation. We characterise the minimum scintillation bandwidth we are sensitive to ($\nu_{\text{min}}$) using simulations as described in Appendix \ref{app:nuMin}, we highlight that this quantity is distinct from the spectral resolution of the data set.

We also fitted for $t_{\text{scatt}}$ in each burst's frequency-integrated pulse profile. By default we assume a scattered Gaussian pulse profile, however, we allow intrinsic burst profiles to comprise multiple Gaussians when necessary. All burst morphology and ACF fitting are performed using a nested sampling technique outlined in \cite{qiu_population_2020}. The frequency evolution of scattering is measured using independent fits to burst sub-bands as done for $\nu_{\text{DC}}$, with $t_{\text{scatt}}$ expected to decrease at higher frequencies. Assuming that $t_{\text{scatt}}$ and $\nu_{\text{DC}}$ evolve in frequency following a power law, we fit for a spectral index describing the evolution of each parameter ($\alpha_{t}$ and $\alpha_\nu$ respectively), in every burst where data permits. In cases where only two sub-bands are used these spectral indices have no measured uncertainty.

We compare our measurements of scintillation or lack thereof with the expected Galactic scintillation ($\nu_{\text{NE2001}}$) using the NE2001 electron density model \citep{cordes_ne2001i_2003}. We note that there can be order of magnitude differences in scatter broadening and scintillation bandwidths for Galactic lines of sight with the same DM \citep{bhat_multifrequency_2004}. Moreover, we use the best fit $\nu_{\text{DC}}$ and $t_{\text{scatt}}$ to compute $C$ as per Eq. \ref{eq:uncertainty}. Finally, when $C\gg1$ we derive the two-screen distance product $L_xL_g$ as expressed in Eq. \ref{eq:twoScreenDist}.

\section{Results}\label{sec:results}
The properties of each burst in our sample can be found in Table \ref{table:burstProperties}, where a `--' denotes parameters that could not be measured or derived. Within our sample, we find three FRBs with convincing evidence of spectral scintillation, from which two-screen constraints can be formed. Of the remaining FRBs, four FRBs were found to contain no spectral scintillation, two contained evidence of spectral structure that could not be confirmed as scintillation and one, FRB\,20191228A, contained instrumental effects for which we could not adequately compensate. In the following sub-sections, we will describe each of these cases in greater detail, with the exception of FRB\,20191228A, which is not analysed further.

\subsection{FRB\,20190608B}\label{subsec:190608}
\begin{figure*}
    \centering
    \includegraphics[width=\textwidth]{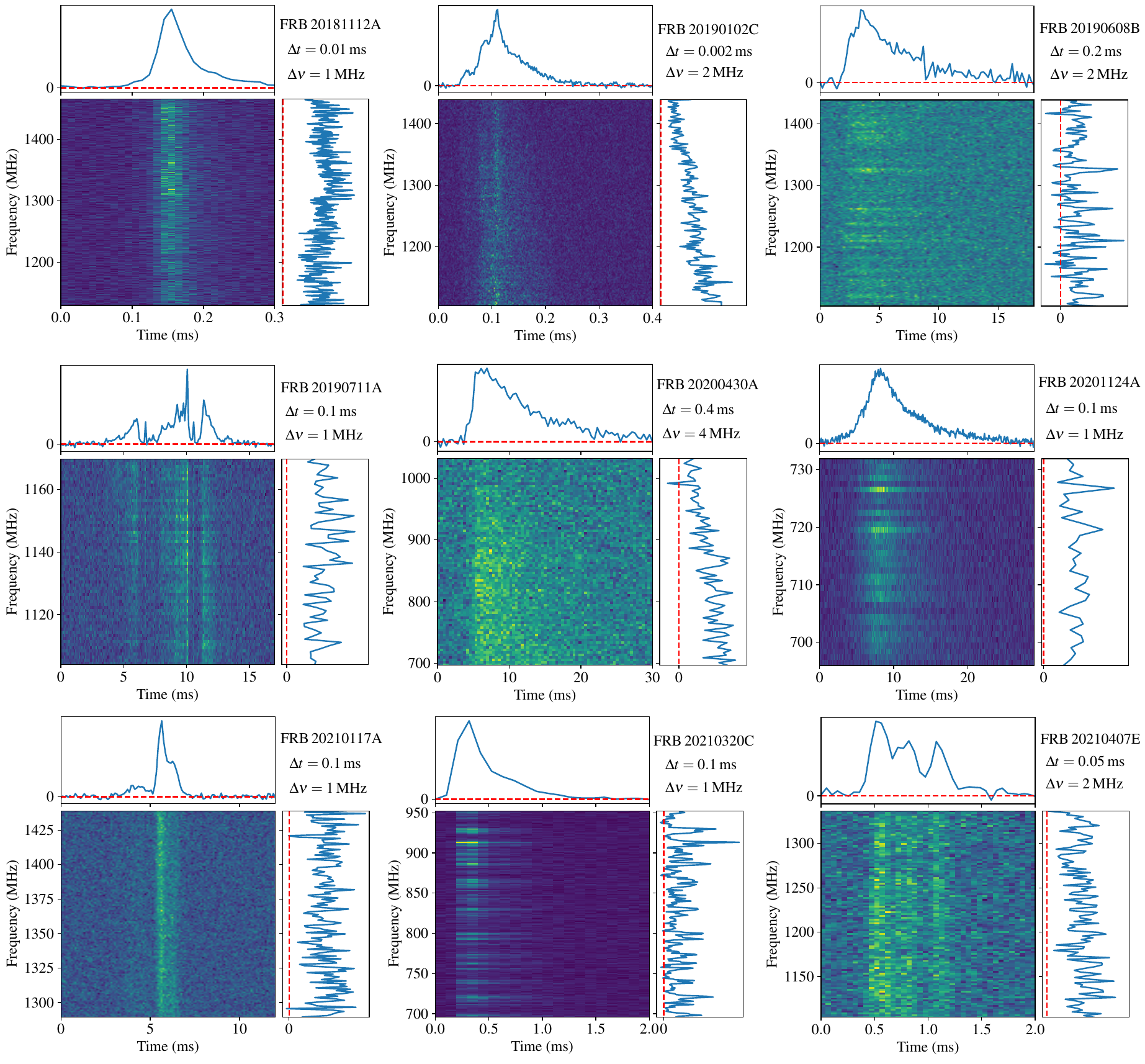}
    \caption{Dedispersed dynamic spectrum of all analysed FRBs. FRB names and spectral and temporal resolutions corresponding to the shown dynamic spectra are labelled in the top right corner of each plot. The top panels of each dynamic spectrum show the burst profiles integrated over frequency, and the right-hand panels are integrated over time.}
    \label{fig:AllDynSpec}
\end{figure*}

\begin{figure}
    \centering
    \includegraphics[width=0.5\textwidth]{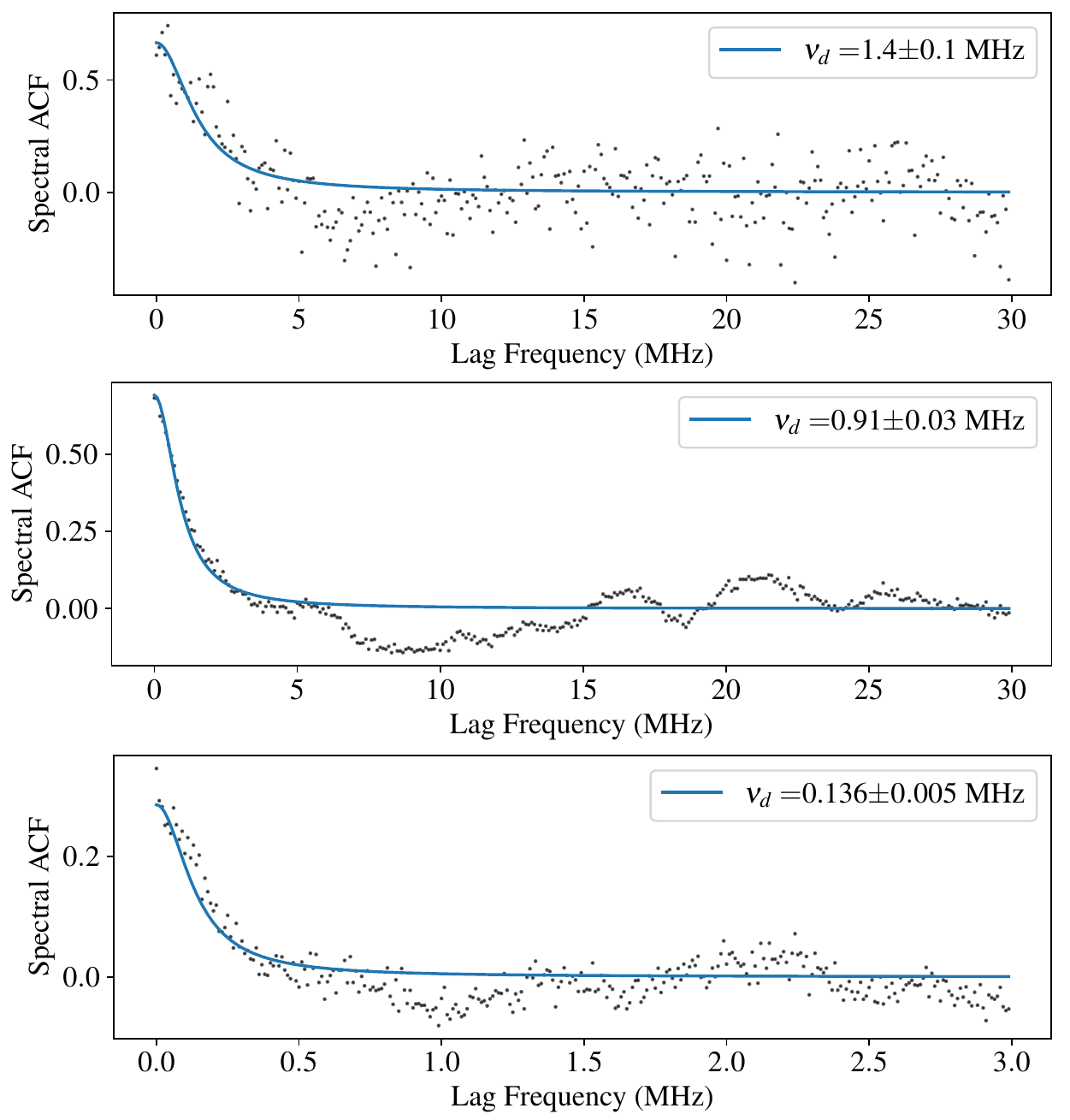}
    \caption{Scintillation fits of FRBs 20190608B, 20210320C and 20201124A from \textit{top} to \textit{bottom}. Black points show the ACF of the time-integrated burst spectra at $0.1$\,MHz, $0.1$\,MHz and $0.01$\,MHz resolution respectively. Blue lines show the best-fit model Lorentzians. The maximum amplitude of the ACF represents the square of the modulation index $m^2$.}
    \label{fig:AllACF}
\end{figure}

\begin{figure}
    \centering
    \includegraphics[width=0.5\textwidth]{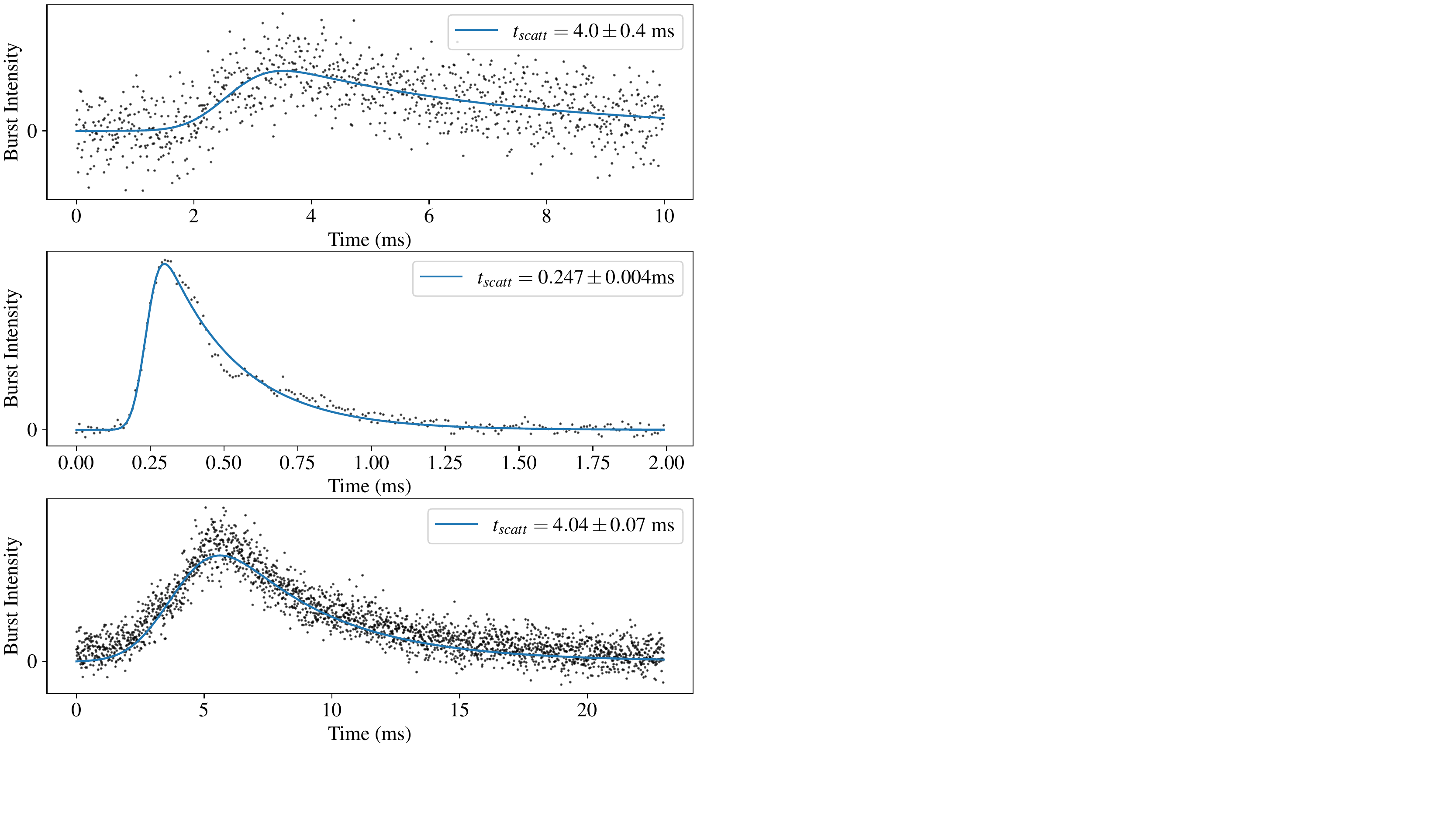}
    \caption{Scattering fits of FRBs 20190608B, 20210320C and 20201124A from \textit{top} to \textit{bottom}. Black points show the frequency-integrated pulse profiles at $10\,\mu$s resolution. The blue lines show the best-fit scattered Gaussian models.}
    \label{fig:AllTScat}
\end{figure}

The dynamic spectrum of FRB\,20190608B is shown in Fig. \ref{fig:AllDynSpec} at a reduced time and frequency resolution of 0.2\,ms and 2\,MHz to improve visual distinction. The burst has the lowest integrated $S/N$ in our sample, however, obvious bands of intensity can still be seen in the dynamic spectrum of the burst. The unnormalised spectrum contains negligible RFI effects and therefore we use all $336$\,MHz of the observed bandwidth centred at $1271.5$\,MHz. Analysing the time-integrated spectrum at $0.1$\,MHz resolution, we measure a high modulation index of $m=0.78$ and $\nu_{\text{DC}}=1.4\pm0.1$\,MHz for the whole band as shown in Fig. \ref{fig:AllACF}. Integrating over frequency, we find a scattering time of $t_{\text{scatt}}=4.0\pm0.4$\,ms, as shown in Fig. \ref{fig:AllTScat}.

Dividing the observation into four subbands we measure the spectral indices of $\nu_{\text{dc}}$ and $t_{\text{scatt}}$ to be $\alpha_\nu=5.8\pm0.5$ and $\alpha_t=-3\pm1$ respectively, as shown in Fig. \ref{fig:AllAlpha}. The frequency evolution of the $t_{\text{scatt}}$ is consistent within $1\sigma$ with $t_{\text{scatt}}\propto\nu^{-4}$ as expected for very strong scattering in a Kolmogorov turbulence with an inner scale \citep{cordes_interstellar_1991, cordes_diffractive_1998}. Conversely, the evolution of $\nu_{\text{DC}}$ is steeper than the Kolmogorov expectation at a marginal significance of $3.6\sigma$.

Given the high modulation index and the positive slope of $\nu_{\text{DC}}$ evolution in frequency, we assume the spectral modulation in this burst is the result of diffractive scintillation of a point-like source. Similarly, the negatively sloped frequency evolution of $t_{\text{scatt}}$ is consistent with multi-path scattering. Combining the measurements of each over the full bandwidth we find $2\pi\nu_{\text{DC}}t_{\text{scatt}}=C\approx35000$, indicating that a single thin screen is insufficient to describe the scattering medium along the line of sight to FRB\,20190608B. Using Eq. \ref{eq:twoScreenDist} we find an upper limit on the two-screen distance product of $L_xL_g\lesssim6\pm1$\,kpc$^2$.  

\subsection{FRB\,20210320C}\label{subsec:210320}
The dynamic spectrum of FRB\,20210320C is shown in Fig. \ref{fig:AllDynSpec}. The burst has $S/N=113$ and a high modulation index of $m=0.83$, consistent with the obvious intensity bands in the burst spectra. The de-dispersion and localisation analysis of this burst will be presented in Shannon et al. (in preparation). Due to the dispersive sweep of $\sim1.8$s across the 336\,MHz ASKAP bandwidth and the $\sim1.6$\,s latency of the detection system, some of the FRB was lost from the voltage buffer before it was downloaded. As a result the burst emission is only found in $257$\,MHz of bandwidth around a central frequency $824.2$\,MHz. The spectral ACF is particularly well fit by a Lorentzian profile with $\nu_{\text{DC}}=0.91\pm0.03$\,MHz, as seen in Fig. \ref{fig:AllACF}. Fig. \ref{fig:AllTScat} shows the best-fit model of the pulse profile, with $t_{\text{scatt}}=0.247\pm0.004$\,ms.

Fitting to four sub-bands, we find $\alpha_\nu=2\pm1$ and $\alpha_t=-3.30\pm0.01$, as per Fig. \ref{fig:AllAlpha} and \ref{fig:AllAlpha} respectively. Each of these parameters evolves with the sign expected for multi-path propagation and are within the ranges observed for pulsars \citep{bhat_multifrequency_2004}. Hence we assume they are caused by scintillation and scattering respectively. Measurements over the whole band yield $C=1410$ with an upper limit on the two-screen distance product of $L_xL_g\lesssim550\pm{30}$\,kpc$^2$.

\subsection{FRB\,20201124A}\label{subsec:201124a}
The dynamic spectrum of FRB\,20201124A is shown in Fig. \ref{fig:AllDynSpec}. The burst appears as a bright narrow-bandwidth pulse, with a $S/N=172$. 

The measured modulation index of the burst is somewhat low at only $m=0.59$, however, the intensity banding in its spectrum motivates us to search for scintillation. To probe the fine spectral structure observed in the burst we analyse the spectrum at $0.01$\,MHz resolution. The Lorentzian structure expected for scintillation provides a good fit to the spectral ACF as plotted in Fig. \ref{fig:AllACF}, with a best fit $\nu_{\text{DC}}=0.136\pm0.005$\,MHz.

\begin{figure}
    \centering
    \includegraphics[width=0.5\textwidth]{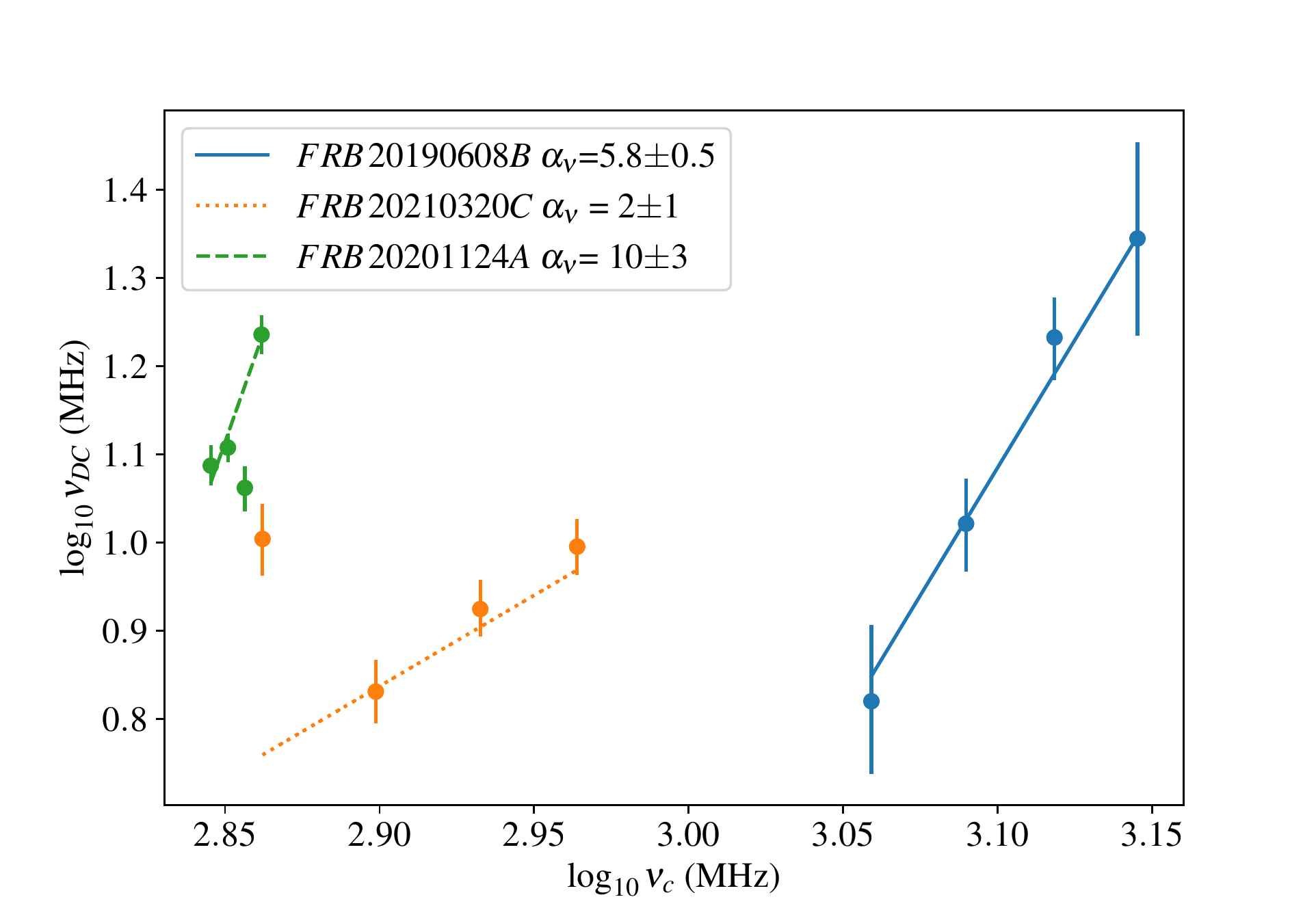}
    \includegraphics[width=0.5\textwidth]{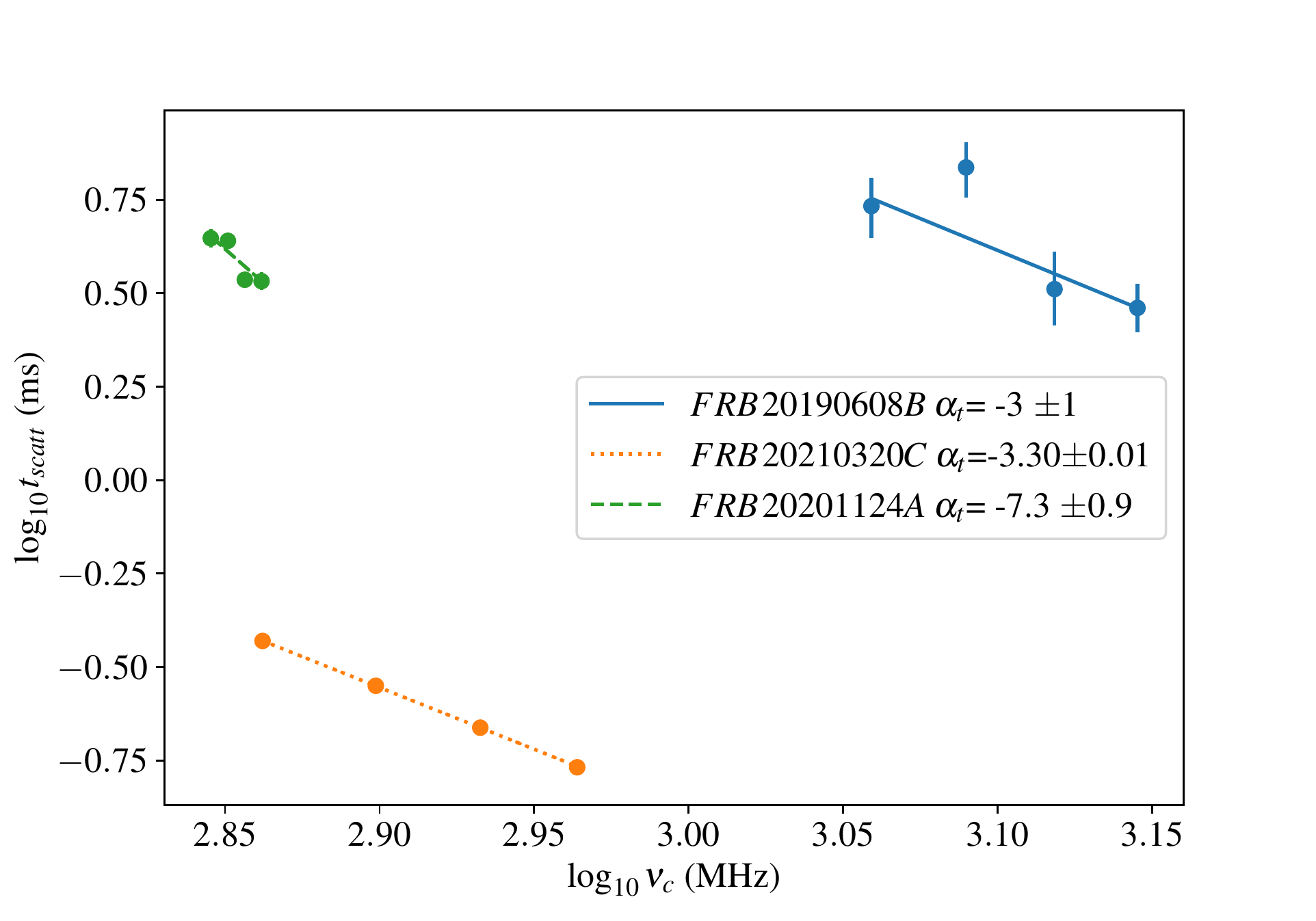}
    \caption{Frequency evolution of $\nu_{\text{DC}}$ (\textit{top}) and $t_{\text{scatt}}$ (\textit{bottom}) modelled from sub-band analysis of each scintillating FRB. The lines show the best-fit power-law models for each case.}
    \label{fig:AllAlpha}
\end{figure}

Despite the narrow bandwidth the burst occupies, its high $S/N$ allows us to measure $\alpha_{\nu}=10\pm3$ across four sub-bands as shown in Fig. \ref{fig:AllAlpha}. This spectral index is consistent with expectations at the $2\sigma$ level. The value of the decorrelation bandwidth is also consistent with the average of other measurements made for FRB\,20201124A \citep{main_scintillation_2021, main_modelling_2022} assuming $\alpha_{\nu}=4$. We, therefore, assume that the frequency structures are caused by scintillation.

We measure the scattering time to be $4.04\pm0.07$\,ms over the whole band, with a $\alpha_t=-7.3\pm0.9$ measured over four sub-bands as shown in Figs. \ref{fig:AllTScat} and \ref{fig:AllAlpha}. This measurement is steeper than the expectation at $3.7\sigma$, however, we note that for this FRB the dynamic range in frequency is extremely limited. Assuming $\alpha_t=-4$, this measurement is consistent with previously measured upper limits on the scattering time for this source \citep{marthi_burst_2022}.

Combined $\nu_{\text{DC}}$ and $t_{\text{scatt}}$ over the used bandwidth yields $C\approx3450$, indicating that a single screen is a poor model for the scattering media along the line of sight. If we assume that the initial scattering screen is unresolved by the first we constrain $L_xL_g\lesssim1.43\pm0.08$\,kpc$^2$, however, we note that in this case, we would expect the observed FRB spectrum to be fully modulated. In \S \ref{sec:discussion} we consider the case of a partially resolved initial scattering screen which could explain the low modulation index.

\subsection{No Observed Scintillation}
For four FRBs within our sample, we observe no spectral scintillation. These are FRBs 20181112A, 20200430A, 20210117A, and 20210407E. As shown in Fig. \ref{fig:AllDynSpec}, the dynamic spectra of these bursts appear spectrally smooth corresponding to relatively constant ACFs, as shown in Fig. \ref{fig:NoScint} contained in the Appendix. As a result, each of these FRBs has a low modulation index, with the exception of FRB\,20200430A which has a modulation index of $m=0.45$, presumably caused by the broad spectral structure in its time-integrated spectrum which we do not attribute to scintillation. Moreover, no significant excess was seen in the spectral ACFs of these bursts at lower resolutions. We are therefore confident in the absence of spectral scintillation on frequency scales above $\nu_\text{min}$ for each of these FRBs, as reported in Table \ref{table:burstProperties}.

\subsection{Anomalous}
We characterise two FRBs within our sample as anomalous. These FRBs, 20190102C and 20190711A, show low spectral modulation indices associated with small excesses in their spectral ACFs. In the case of FRB\,20190102C the ACF of the whole band shows a broad spectral structure which we do not associate with scintillation, and a sharper ACF peak at low spectral lags ($<5$\,MHz), as shown in Fig. \ref{fig:Strange190102}, which is potentially consistent with scintillation. We fail to find an ACF excess when we decompose the burst into four sub-bands, however, reducing the division to two sub-bands yields a reasonable fit as shown in Fig. \ref{fig:Strange190102}. Derived $\nu_{\text{DC}}$ values evolve in the expected direction for scintillation, however, given the low number of sub-bands we are unable to estimate the error on the spectral index of $\alpha_\nu=10$. 

In the case of FRB\,20190711A the ACF is well fit by the expected lorentzian form of scintillation, as shown in Fig. \ref{fig:Strange190711}. Analysis of the sub-bands measures $\alpha_\nu = -10\pm5$, contrary to the expectation for scintillation decorrelation bandwidths to increase in size with frequency. We highlight however the low fractional bandwidth ($\approx0.06$) over which these data are measured.

Both FRBs show complex pulse profiles, with multiple components. Owing to the computational load associated with multi-component fitting we only model the temporal properties of each of these bursts with two sub-bands. In each case the short timescale structure of each of these bursts could also cause intrinsic spectral structure on the reciprocal scale \citep{nimmo_burst_2022}. Given the uncertainty associated with their measurements and the low $C$ values, which can indicate the consistency of spectral structures with self-noise, we conclude that there is insufficient evidence to prove scintillation in these cases.

\section{Discussion}\label{sec:discussion}

For the cases where we find convincing evidence for scintillation and pulse broadening, i.e. FRBs 20190608B, 20201124A, and 20210320C, the scattering geometry is constrained by the $L_xL_g$ product upper limit. FRBs 20190608B and 20201124A provide particularly tight constraints.

\subsection{FRB\,20190608B}\label{subsec:190608Dis}

Due to limitations in spectral resolution, the presence of diffractive scintillation was unable to be confirmed in a previous analysis of FRB\,20190608B \citep{day_high_2020}. Without the presence of this scintillation the position of the screen causing temporal scattering in the burst had to be inferred indirectly from estimates of the host galaxy properties \citep{chittidi_dissecting_2021} and the properties of the cosmic web along the FRB line of sight \citep{simha_disentangling_2020}. The joint conclusion of these studies is that the temporal scattering in FRB\,20190608B is likely contributed by a region within the host galaxy as there are no cosmic web structures or foreground galaxies intersecting the line of sight sufficiently to explain the large scattering time. 

By confirming scintillation and placing an upper limit of $L_xL_g\lesssim6\pm1$\,kpc$^2$ our results provide a direct constraint on the scattering geometry. For similar values $L_x\simeq L_g$, the screens must be contained within the host and Milky Way galaxies respectively. 

\begin{figure}
    \centering
    \includegraphics[width=0.5\textwidth]{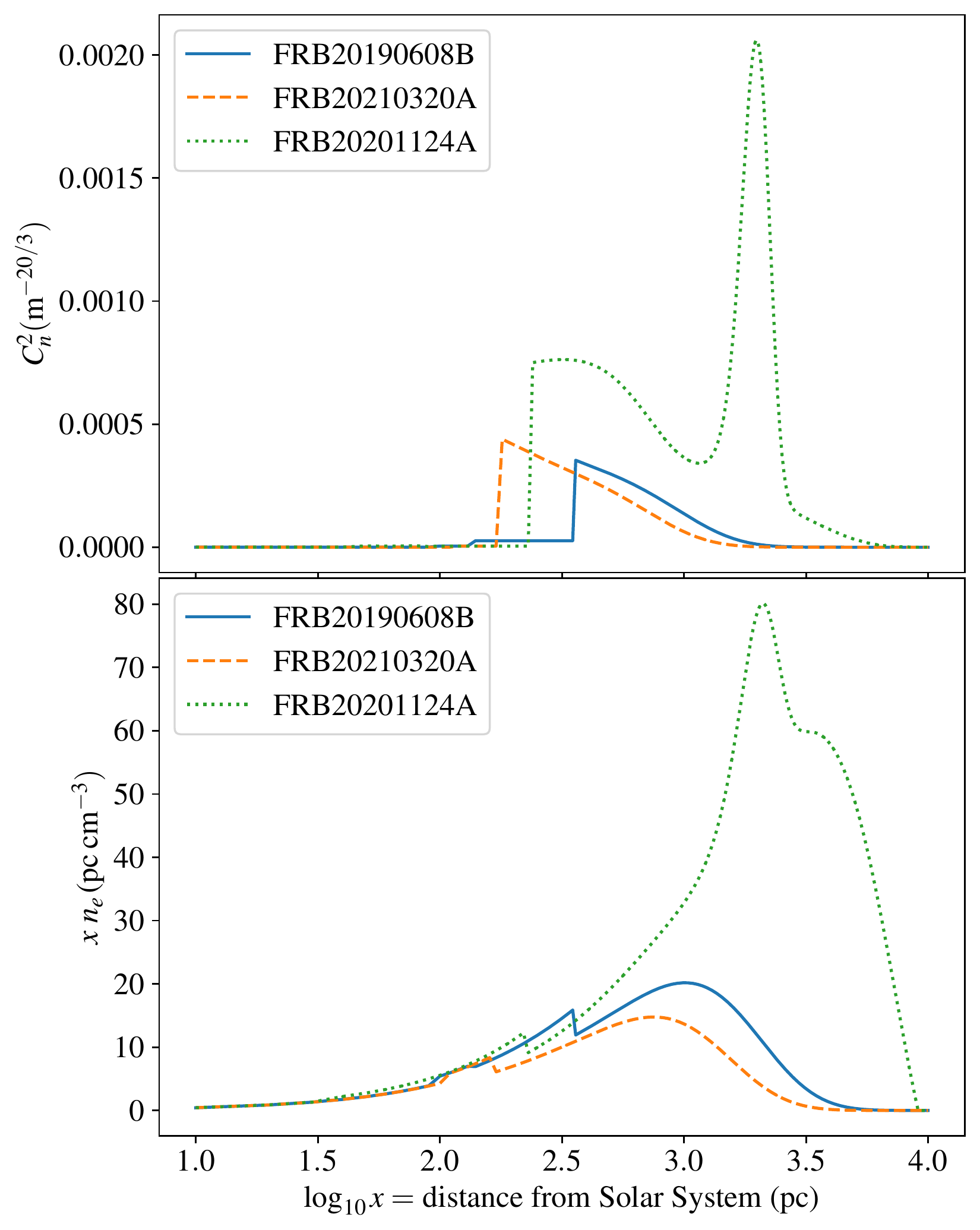}
    \caption{\textit{Top}: The expected differential scattering measure contributed by the Milky Way for scintillating FRB lines of sight as calculated from the NE2001 model \citep{cordes_ne2001i_2003}. \textit{Bottom}: For the same lines of sight, the expected contribution to DM as a function of $\log_{10}x$ where $x$ is the distance from the Solar system. For each FRB line of sight the maximum of $C_n^2$ occurs at $0.36$\,kpc, $0.18$\,kpc, and $1.98$\,kpc respectively, corresponding to sharp changes in DM contribution.}
    \label{fig:FRBneCn2}
\end{figure}

By measuring the angular broadening extent of an FRB using VLBI (very long baseline interferometry), the effective distance to the relevant scattering screen can be determined \citep{ocker_constraining_2021}. This has been done for FRB\,20121102 using the European VLBI network \citep{marcote_repeating_2017}. The effective distance to its Galactic scattering screen is constrained to be consistent with a peak in differential scattering measure associated with a sharp change in electron density predicted by the NE2001 model \cite{ocker_constraining_2021}. Fig. \ref{fig:FRBneCn2} shows the $C_n^2$ and differential DM estimated for each of our scintillating FRB lines of sight. Using the peak in $C_n^2$, corresponding to a sharp change in differential DM, we estimate $L_g\approx0.36$\,kpc , corresponding to $L_x\lesssim16.7$\,kpc. This region corresponds to the host galaxy of FRB\,20190608B and therefore our direct constraints support the conclusions of \cite{simha_disentangling_2020, chittidi_dissecting_2021}. Additionally, by assuming $L_g\approx0.36$\,kpc, $\nu_{\text{DC}}$ and the fully modulate version of Eq. \ref{eq:generalTwoScreen} can be used to constrain the product 
\begin{equation}\label{eq:IGMConstraint}
    t_{\text{scatt}}(1+z_d)\frac{D_{ds,x}}{D_{d,x}}\lesssim \frac{\nu_{\text{DC}}D_s}{2\pi\nu^2L_g}
\end{equation}
where $D_{d,x}$ is the distance to the extragalactic scattering screen at redshift $z_d$, and $D_{ds,x}$ is the distance between the screen and the host. From this constraint, we place an upper limit on the amount of scattering caused by the IGM. In the case of FRB\,20190608B, we find $t_{\text{scatt}}(1+z_d)D_{ds,x}/D_{d,x}\lesssim1.7\times10^{-7}$\,s, corresponding to less than $0.43\,\mu$s of scattering at $1$\,GHz (assuming a $\nu^{-4}$ scaling), for a screen at redshift $z\approx0.056$, where $D_{ds,x}/D_{d,s}\approx1$ and the scattering time associated with a given scattering measure is maximised \citep{macquart_temporal_2013}.

\subsection{FRB\,20210320C}\label{subsec:210320Dis}

FRB\,20210320C provides perhaps our sample's best example of scattering and scintillation, with the burst morphologies presented in Fig. \ref{fig:AllACF} and Fig. \ref{fig:AllTScat} showing good agreement with the expected shapes and frequency evolutions for diffractive scintillation and pulse broadening. The small amount of observed scattering in this case, however, results in only a loose constraint on the scattering geometry of $L_xL_g\lesssim550\pm30$\,kpc$^2$. From Fig. \ref{fig:FRBneCn2} the peak in turbulence strength is close to the observer at a distance of $0.18$\,kpc, which corresponds to $L_x\lesssim3000$\,kpc. As such, the scattering cannot be definitively constrained to the host galaxy. We note however, that the observed scattering must still be occurring within the first $\approx0.3\%$ of the total path length from the host and so cannot be due to some diffuse component of the IGM as its contribution to the scattering would characteristically peak halfway between the source and the observer. For IGM scattering in general, by assuming $L_g\approx0.18$\,kpc, we find $t_{\text{scatt}}D_{ds,x}/D_{d,x}\lesssim1.1\times10^{-6}$\,s, using Eq. \ref{eq:IGMConstraint}. This corresponds to less than $0.44\,\mu$s of scattering at $1$\,GHz (assuming a $\nu^{-4}$ scaling), for a screen at redshift $z\approx0.126$, where $D_{ds,x}/D_{d,s}\approx1$.

The host galaxy localisation image of FRB\,20210320C, shows a faint object nearby to the line of sight. The redshift of this object has yet to be determined, however, if it lies foreground to the host galaxy at a similar redshift it may be the source of the observed scatter broadening.

\subsection{FRB\,20201124A}
Conversely to the other scintillating FRBs in our sample, FRB\,20201124A is a closely studied repeating FRB with existing measurements for its scattering time and decorrelation bandwidth. Analysis by \citet{main_scintillation_2021} measured $\nu_{\text{DC}}\approx0.1$\,MHz and $t_{\text{scatt}}\approx11$\,ms at a central frequency of 575\,MHz. Substituting these values into Eq. \ref{eq:twoScreenDist} yields $L_xL_g\lesssim 0.6$\,kpc$^2$, which is tighter than the limit we derive, $L_xL_g\lesssim1.43\pm0.08$\,kpc$^2$, consistent with the expected steep frequency dependence of the constraints \citep{main_scintillation_2021}.

\begin{figure}
    \centering
    \includegraphics[width=0.5\textwidth]{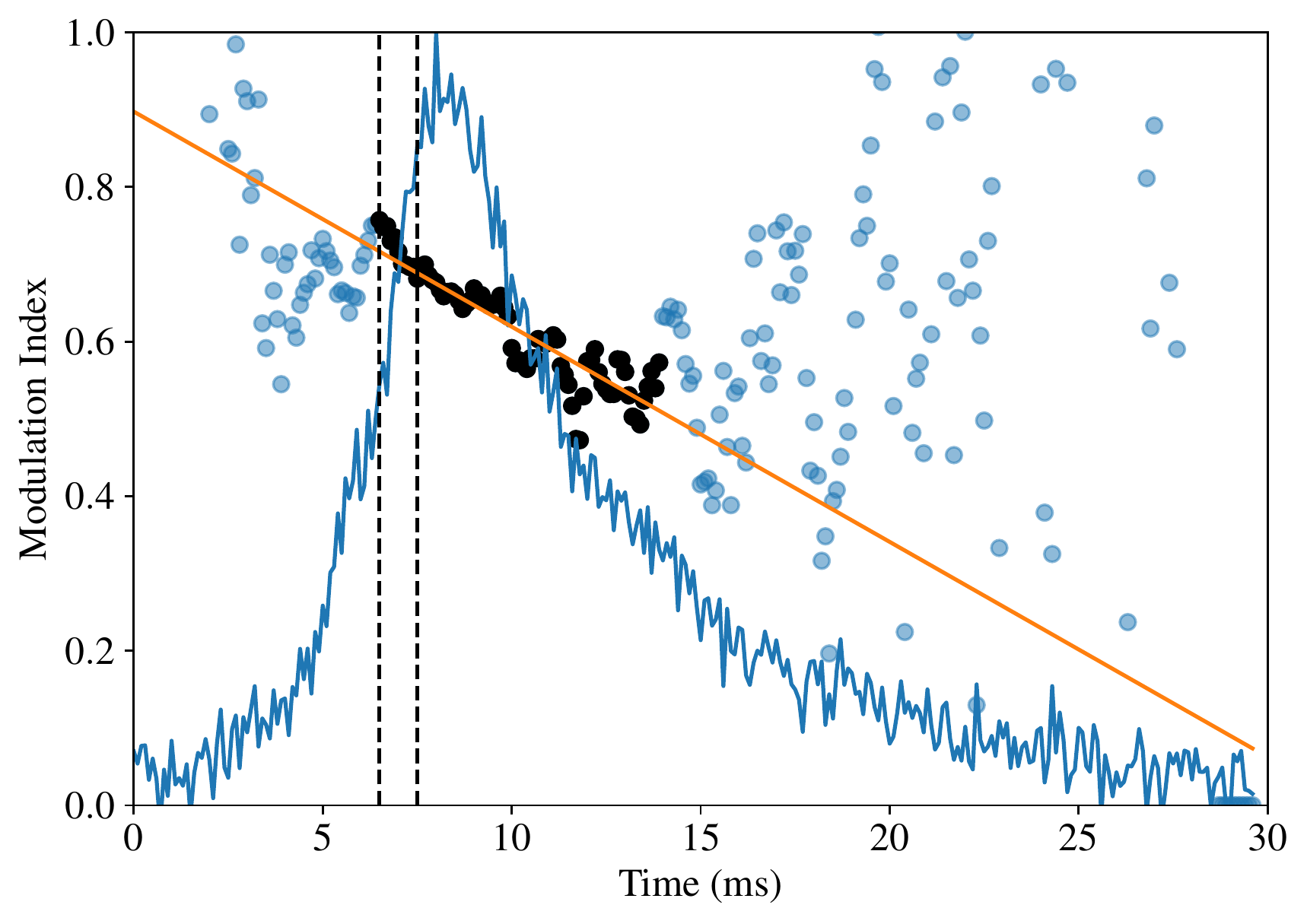}
    \caption{Modulation index of FRB\,20201124A as a function of time. Scatter points show the modulation index calculated for the burst spectrum at a resolution of $0.01$\,MHz integrated over $1$\,ms of the bursts time profile (shown by the blue line) beginning at the time marked by the point. The dotted black lines depict the area of integration in time. The linear model of modulation index decay is shown by the orange line. The points used in the fit are shown in black.}
    \label{fig:20201124AModEvo}
\end{figure}

Despite the evidence for scintillation, the observed modulation index of FRB\,20201124A remains too low to be consistent with the full modulation expected for diffractive scintillation of a point-like source. In this context, a source will be considered point-like if it satisfies Eq. \ref{eq:coherenceLengthCond}. If the equation is violated, we enter the regime of diffractive scintillation of an extended source. Here, the modulation index of the spectral scintillation will begin to decrease as the angular extent of the scattering disk of the first screen increases \citep{narayan_physics_1992}. Within the temporal profile of scattered bursts, later times are associated with larger angular extents in the scattering disk. Similarly to the analysis of \cite{masui_dense_2015}, we can analyse the modulation index of the burst as a function of time to identify whether the entire angular extent of the scattered image undergoes the same scintillation. If the scattered image associated with the observed temporal broadening is responsible for the suppression of Galactic scintillation, we expect that the later parts of the burst, with larger angular extents, will show lower modulation indices. Fig. \ref{fig:20201124AModEvo} shows the evolution of the modulation index over the duration of the burst in increments of $0.1\,$ms. At each increment a $1$\,ms wide boxcar of the burst's dynamic spectrum is used to calculate the modulation index, effectively smoothing the result to boost $S/N$. As seen in Fig. \ref{fig:20201124AModEvo} the modulation index shows a small decrease of $\sim0.2$ over the main component of the burst profile with a large variance in $m_g$ displayed on either side of the burst as $S/N$ decreases. A linear model shows reasonable agreement with the data, as would be expected for a circularly symmetric scattered image, where separation in time is linearly proportional to angular offset. The low modulation index of spectral scintillation in FRB\,20201124A may, therefore, indicate that the scattering screen at the host is partially resolved by the Milky Way scattering screen. 

For the fully modulated case, the coherence length of a wave incident on the second screen can only be constrained to be larger than the projected scattering angle length. In the partially resolved case, however, it can be solved for exactly using the modulation index. This, in turn, allows the two-screen distance product to be specified exactly, as, (see \ref{app:derivation} for derivation)
\begin{equation}\label{eq:PartiallyResolved2ScreenScat}
L_xL_g \approx \frac{D_{s}^2}{2\pi\nu^2(1+z)\,m_g^2}\frac{\nu_{DC}}{t_{\text{scatt}}},
\end{equation}
where $m_g$ is the modulation index of the Galactic screen. Solving this for the case of FRB\,20201124A indicates that the two-screen distance product should be equal to $L_xL_g \approx 4$\,kpc$^2$.

A recent study of the annual variation of scintillation in FRB\,20201124A has revealed that the scattering screen contributing the observed Galactic scintillation is much closer than the peak in $C_n^2$ at $\sim2$\,kpc suggests, located at around $L_g=0.40-0.46$\,kpc, depending on the isotropy of the screen \citep[][]{main_modelling_2022}. We discuss the potential impact of screen anisotropy in appendix \ref{sec:anisotropy}. Taking the case of the uniform two-dimensional screen, $L_g=0.46$\,kpc, we can approximate the distance between the source and host screen to be $L_x\approx9$\,kpc, which is greater than the optical extent of the host galaxy \citep{xu_fast_2022}. This indicates that, if the angular broadening associated with the measured scattering tail is suppressing the observed Galactic scintillation, that scattering is likely occurring in the halo of the galaxy, rather than in the circum-burst environment or the host ISM. 

The low modulation index of the burst could also be caused by angular broadening from a third screen along the line of sight which contributes negligibly to the observed scattering and scintillation of the burst. This is precisely the inverse of the case discussed in \S \ref{subsec:190608Dis} regarding limits on IGM scattering. The two possible locations for this potential third screen are within the Milky Way or the IGM. The case where the third screen is also within the host galaxy is captured implicitly by the above discussion. Already, some motivation for a third screen within the Milky Way exists, in the form of the peak in $C_n^2$ at $\approx2\,$kpc shown in Fig. \ref{fig:FRBneCn2}, which we know is not associated with the observed scintillation. The scattering time required from the third screen to reduce the modulation index of the Galactic scintillation is given by 
\begin{equation}
    t_{\text{scatt}}\approx \frac{\nu_{\text{DC}}D_{d,x}}{2\pi\nu^2m_g^2L_g},
\end{equation}
adapted from Eq. \ref{eq:IGMConstraint}. Solving for $m_g=0.59$ yields $t_\text{scatt}\approx1.3\times10^{-7}\,\mu$s at $1\,$GHz (assuming a $\nu^{-4}$ scaling), showing that the foreground Galactic scintillation can be suppressed with very little additional scattering from a third screen in the Milky Way. From this result, we conclude that while it is possible for such a scenario to be true, it is more likely that any third screen in the Milky Way would completely suppress the scintillation from the foreground screen at $0.46\,$kpc and is therefore inconsistent with our observations. 

If the third screen is instead placed within the IGM, we can use the inverse of Eq. \ref{eq:IGMConstraint}, dividing the right-hand side by $m_g^2$ to take the partial modulation into account. This yields $t_{\text{scatt}}(1+z_d)D_{ds,x}/D_{d,x}\approx 0.027\,\mu$s at $1\,$GHz (assuming a $\nu^{-4}$ scaling), we plot this result as a function of $D_{d,x}$ in Fig. \ref{fig:20201124AIGM}. The scattering times required to cause $m_g=0.59$ are reasonable expectations for scattering from the IGM \citep{macquart_temporal_2013} and would be invisible in the temporal profile of the burst. The range of decorrelation bandwidths corresponding to the spectral scintillation also imposed by an IGM screen, however, falls mostly within our detectable range and therefore should appear in our observations. As a result we find it unlikely that angular broadening from a third screen in the IGM can adequately explain the observed scintillation modulation. We note however that for IGM screens closer than $50\,$Mpc the decorrelation bandwidth would be greater than our observed bandwidth, and therefore undetectable. 

Given the issues outlined above with a third screen interpretation, coupled with the observed evolution of the modulation index over the burst, we tentatively conclude that the most likely scenario is that the Galactic scintillation observed in the burst is suppressed by the angular broadening corresponding to the observed temporal broadening. As such FRB\,20201124A is a potential candidate of interest for probing the CGM, and we recommend its modulation index and scattering times be studied in detail in future statistical studies of its repeating bursts.

\subsection{Circum-burst Scattering}
Given the localisation of the scattering to within 0.4\,AU of the source for FRB\,20190520B \citep{ocker_scattering_2022}, it is prudent to consider the ramifications if this were typical for all FRBs. The extremely low value of $L_x$ in each case would leave $L_g\lesssim D_s$, and hence the position of the screen responsible for the spectral scintillation would be unbounded. While the diffuse IGM is not expected to cause sufficient scattering to account for FRB temporal broadening, it is expected to be able to cause the microsecond level scattering required to see scintillation on megahertz scales \citep{macquart_temporal_2013}. It is, therefore, possible, if the screen causing the observed temporal broadening of FRBs is associated with the circum-burst environment, that the observed spectral scintillation could come from the IGM, invalidating the previous IGM scattering constraints in \S \ref{subsec:190608Dis} and \ref{subsec:210320Dis}. However, in order for no additional scintillation from the Milky Way to be observed, consistent with our observations, which show only one scale of frequency modulation, the angular broadening from the IGM must be such that any subsequent Milky Way scintillation is suppressed. 

The two-screen interaction between the IGM and the Milky Way, can be considered using Eq. \ref{eq:IGMConstraint} where $t_{\text{scatt}}=1/2\pi\nu_{\text{DC}}$ and $D_{ds,x}=D_{d,x}$, and assuming that the Milky Way scintillates as expected by NE2001, Galactic scintillation will be suppressed for all
\begin{equation}\label{eq:IGMMWSupp}
    L_g\gtrsim \frac{D_s}{\nu^2(1+z_d)}\nu_{\text{DC}}\nu_{\text{NE2001}},
\end{equation}
where $z_d$ is the redshift of the IGM screen. For FRBs 20190608B and 20210320C, Eq. \ref{eq:IGMMWSupp} yields $L_g\gtrsim1$\,kpc. Fig. \ref{fig:FRBneCn2} shows that for both FRBs, the Galactic scattering screens are expected to be closer than 1\,kpc, and therefore would still cause visible scintillation in each, in addition to the IGM scintillation. This is inconsistent with our observations, and therefore we find it unlikely that the observed scintillation comes from the IGM. This agrees with the observed correlation between FRB scintillation and expectations from Galactic electron distribution models \citep{schoen_scintillation_2021}. We highlight, however, that scintillation from the IGM is a reasonable possibility for FRBs that appear sufficiently point-like. To investigate this possibility further we recommend a statistical study that measures the correlation between FRB redshift and $\nu_{\text{DC}}$. The use of redshift is preferable to dispersion measure as the host contribution to dispersion is difficult to separate from the IGM contribution. If a significant fraction of FRBs contain scintillation from the IGM, we expect that an anti-correlation between redshift and $\nu_{\text{DC}}$ will be present.

\subsection{Galactic Scintillation}\label{subsec:galacticScint}
Galactic electron distribution models such as NE2001 and YMW16 are widely used to determine the expected Galactic scintillation along a given line of sight. While a correlation between observed scintillation and model expectations has been established \citep{schoen_scintillation_2021}, it has also been shown that Galactic scintillation can be dominated by extremely small scale features \citep{stinebring_scintillation_2006, trang_modeling_2007, brisken_100_2010}. Existing models are therefore not expected to satisfactorily map the distribution of Galactic scintillation, due to their limited number of components \citep{yang_dispersion_2017}. FRB\,20201124A is a prime example of this, demonstrating significantly less scattering than expected for its line of sight \citep{main_modelling_2022, main_scintillation_2021}. Motivated by this we compare our measured Galactic scintillation to expectations.

\begin{figure}
    \centering
    \includegraphics[width=\textwidth, angle=-90, origin=c]{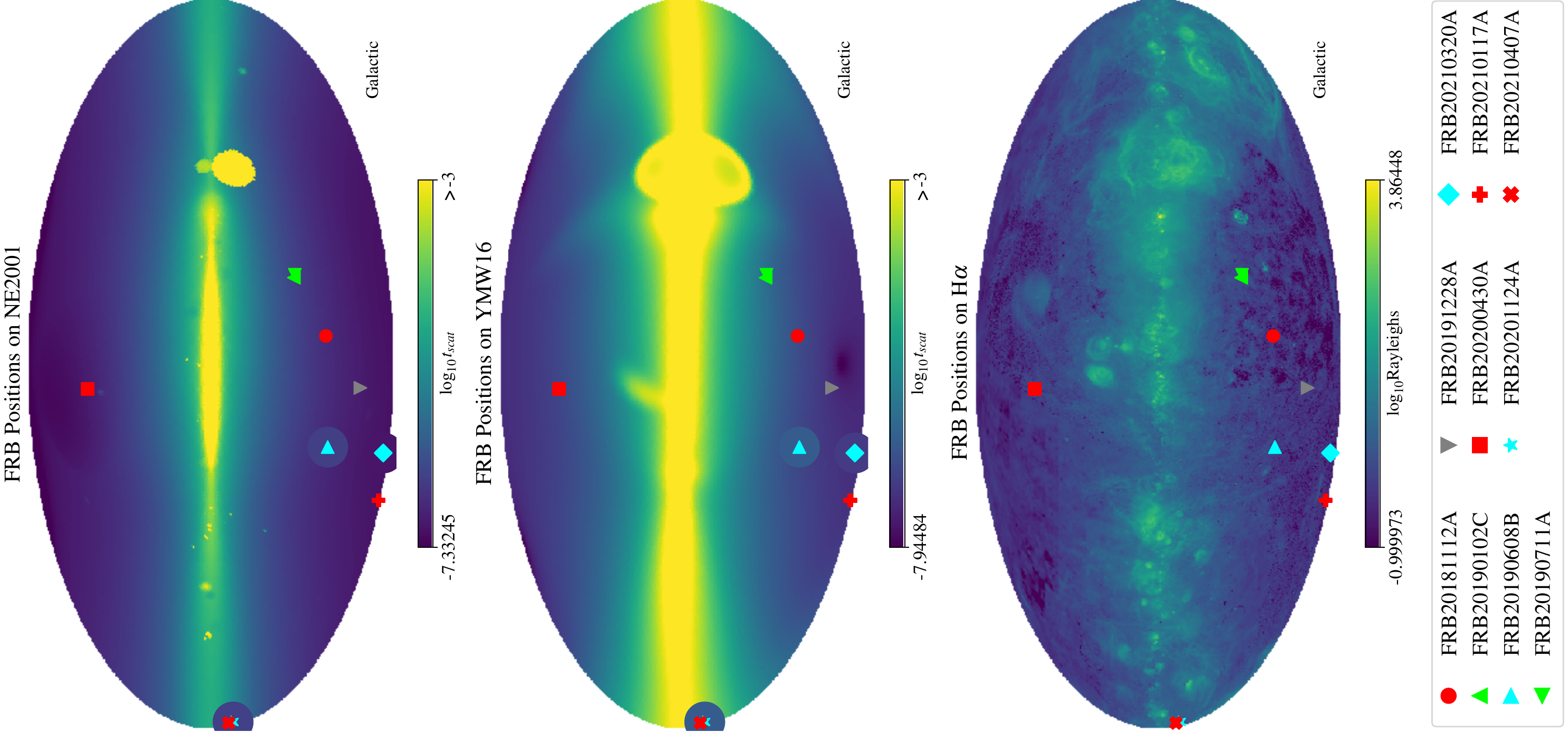}
    \caption{FRB localisations superimposed on Galactic electron distribution models and measured H$\alpha$ intensities. Cyan points represent FRBs confirmed to scintillate. Red points represent FRBs confirmed to not scintillate. Green points represent FRBs where evidence of scintillation was inconclusive. The grey point represents FRB\,20191228A, where data was corrupted. Around each cyan point on the Galactic electron density maps (\textit{top} and \textit{middle}) we change the region colour to represent the FRBs measured Galactic scintillation time for comparison with the model estimates. To aid visual distinction scattering times greater than or equal to $10^{-3}$\,s are shown as the same colour. \textit{Top}: Ne2001 model \citep{cordes_ne2001i_2003}. \textit{Middle}: YMW16 model, which uses DM and the Bhat relationship to predict scattering \citep{bhat_multifrequency_2004, yang_dispersion_2017}. \textit{Bottom}: The \citet{finkbeiner_full-sky_2003} H$\alpha$ all-sky intensity map.}
    \label{fig:AllSkyGEDM}
\end{figure}

We plot the scintillation measured for FRBs in our sample against the expected Galactic scattering from these models in Fig. \ref{fig:AllSkyGEDM}, scaling all measures to 1\,GHz using a $\nu^{4}$ scaling relation. While our observations of FRB\,20201124A also show anomalously low scattering ($\sim1/20$\,NE2001 and $\sim1/1000$\,YMW16), consistent with previous observations \citep{main_scintillation_2021}, scintillation in FRBs 20190608B and 20210320C agree with estimates from both YMW16 and NE2001 to within a factor of $\approx2$. FRBs 20190608B and 20201124A show very similar levels, however, FRB\,20210320C has a much lower scattering time as expected for higher Galactic latitudes.

The four non-scintillating FRBs within our sample are distributed over a range of Galactic latitudes where we have observed FRB scintillation, as shown in Fig. \ref{fig:AllSkyGEDM}. Given the smoothness of the expected Galactic electron distributions, we expect to observe similar Galactic scintillation in each of these cases. The lack of apparent scintillation could suggest that either the Galactic electron distribution models fail to capture coarse variations in the expected scattering or that angular broadening from another screen is suppressing Galactic scintillation. For the three non-scintillating FRBs with confirmed redshifts, we can use Eq. \ref{eq:twoScreenDist} to calculate the distance an extragalactic scattering screen would need to be from the host, in order to suppress the expected Galactic scintillation. Using NE2001 we find that, in each case, the screen causing the temporal broadening of these FRBs would need to be greater than $\sim10$\,kpc away from the source (in some cases several Mpc further) in order to suppress the expected Galactic scintillation.

Scintillation could also be suppressed by a third scattering screen contained within the IGM. In this case Eq. \ref{eq:IGMConstraint} can be inverted to determine the minimum value of $t_{\text{scatt}}(1+z_d)D_{ds,x}/D_{d,x}$ for an IGM screen to begin suppressing the expected Galactic scintillation. For FRBs 20181112A, 20200430A and 20210117A, respectively, we find that the IGM would need to cause scattering well in excess\footnote{For scattering times, exactly equal to the limit the modulation index will be one, and will decrease as $\sqrt{t_{\text{scatt}}/t_{\text{limit}}}$} of $3.3\,\mu$s, $0.11\,\mu$s and $5.0\,\mu$s at $1\,$GHz (assuming a $\nu^{-4}$ scaling) for screens at redshifts 0.198, 0.076 and 0.099 (where $D_{ds,x}/D_{d,x}\approx1$, resulting in the maximum scattering time for a given scattering measure). 

From \cite{macquart_temporal_2013}, observed scattering times are related to the effective scattering measure (SM$_{\text{eff}}$) via
\begin{equation}
    t_{\text{scatt}}(1+z_d)\propto \left(\nu^{-4}\right)\left(\frac{D_{d,x}D_{ds,x}}{D_s}\right)\left(\text{SM}_{\text{eff}}\right),
\end{equation}
for scattering dominated by diffractive scales below a constant inner turbulence scale. 
Using the redshifts associated with each source and midpoint, the limits on the IGM scattering times derived for both scintillating and non-scintillating FRBs can then be converted to limits proportional to SM$_{\text{eff}}$. For the case of a diffuse IGM, \cite{macquart_temporal_2013} show that SM$_{\text{eff}}$ should be a monotonically increasing function of redshift, which is inconsistent with the limits we derive, as shown in Fig. \ref{fig:AllSMConst}. As a result, a diffuse IGM component cannot be responsible for suppressing Galactic scintillation expected from NE2001 in our non-scintillating bursts given our observations of scintillation in others. We, therefore, suggest that it is unlikely that angular broadening associated with scattering from the diffuse IGM, or regions within the host galaxy of each burst is suppressing the Galactic scintillation of the non-scintillating FRBs in our sample. Rather, it is more likely that either 1) the true scintillation bandwidths are different from that predicted under the NE2001 and YMW16 models, similar to the case of FRB\,20201124A, or 2) significant structures in the IGM, such as a galaxy halo, are intervening between the source and the observer, resulting in significant angular broadening. In the case of FRB\,20181112A such a structure exists in the form of a foreground galaxy intervening at $z=0.36738$ \citep{prochaska_low_2019}. However, at this redshift a screen contributing the maximum amount of scattering allowed for this burst ($0.0278\,$,ms, as per Table \ref{table:burstProperties}) results in a modulation index of $m_g=0.46$ and should be visible in its ACF (see Fig. \ref{fig:NoScint}).

As demonstrated in \citet{morgan_census_2022}, there exists a strong correlation between the angular broadening of extragalactic point sources and the intensity of Galactic H$\alpha$ emissions. Angular broadening of extragalactic sources is weighted approximately uniformly with distance for Galactic scattering screens \citep{cordes_ne2001i_2003}, and therefore, ionised regions, shown by Galactic H$\alpha$ observations, are expected to contribute significantly to angular broadening by the ISM. Conversely, the scintillation bandwidths and scattering times associated with the ISM are weighted more heavily towards the middle of the path length \citep{cordes_ne2001i_2003}, allowing background regions of lower H$\alpha$ intensity to cause greater scattering times than more intense foreground regions, confusing any correlation. Moreover, close to the Galactic plane the ionised region responsible for the observed scattering may not be visible in H$\alpha$, due to extinction \citep{finkbeiner_full-sky_2003}. As such, we do not necessarily expect to observe a strong correlation between FRB $\nu_{\text{DC}}$ and H$\alpha$. The association of scintillation in FRB\,20201124A to a more local screen, however, does provide some motivation to search for scintillation screens locally, where extinction should be relatively low. Thus, we also compare scintillation in our sample to a Galactic map of H$\alpha$ intensity \citep{finkbeiner_full-sky_2003} as shown in Fig. \ref{fig:AllSkyGEDM}. We find no obvious relation between scintillating and non-scintillating FRBs and H$\alpha$ intensity or variance in $2.5^{\circ}\times2.5^{\circ}$ area surrounding each FRB line of sight. We note, however, that the size of a reasonable scattering disk lies well below the resolution of data used to compose this map and therefore may show correlation with smaller scale H$\alpha$ structures.

\section{Conclusion}\label{sec:conclusion}
The location of the dominant scattering screens contributing to the temporal broadening and spectral scintillation of FRBs has important ramifications for many areas of astrophysics. Scattering near the host galaxy or circum-burst environments, such as FRB\,20190520B affects our understanding of progenitor evolution; scattering in intervening galaxies could constrain the presence of cold cloudlets in the CGM; finally, scattering in our own Galaxy could inform models of the Galactic electron distribution. For apparently non-repeating FRBs scattering can be difficult to localise, but if bursts are observed to scatter and scintillate independently then a two-screen model can be used to make direct constraints. 

In this work, we have measured the level of scattering and scintillation in 10 CRAFT FRBs with high spectro-temporal resolution and applied the two-screen model developed by \citet{masui_dense_2015} and \citet{ocker_large_2022} to place constraints on the distances to their respective scattering screens. We find strong evidence for scattering and scintillation in three FRBs, and strong evidence for no spectral modulation in four FRBs. The remaining are indeterminate. Of the scintillating FRBs the scattering in FRB\,20190608B is robustly associated with the host galaxy in agreement with previous estimates; the scattering in FRB\,20210320C must occur within 3\,Mpc of the host; finally, we find that the scattering in FRB\,20201124A is likely associated with its host galaxy environment, however, the low modulation index of its Galactic scintillation suggests the dominant scattering region may be in the halo rather than the host ISM. The Galactic scintillation of FRBs 20190608B and 20210320C are in general agreement with the scintillation expected from Galactic models YMW16 \citep{yang_dispersion_2017} and NE2001 \citep{cordes_ne2001i_2003}. However, the anomalously low scattering of FRB\,20201124A and the definitive lack of scintillation in four FRBs indicates that, if the observed pulse broadening is contributed by host galaxy ISMs or circum-burst environments, existing models may be poor estimators of the scattering times associated with the Milky Way's ISM, as has been noted already by \citet{ocker_constraining_2021} in the case of YMW16. Additionally, we find no obvious relationship with the large-scale mean and variance of surrounding Galactic H$\alpha$ emission. We leave a statistical comparison of scintillation quantities with other burst and host galaxy properties to a future study, once the sample of high-resolution bursts has been expanded.

With the automated CELEBI post-processing pipeline now operational and the CRAFT Coherent upgrade expected soon we expect that the number of observed scintillating FRBs will grow, allowing for a statistical study of their Galactic and extragalactic screen properties. Furthermore, we highlight that targeting low Galactic latitudes for FRB searches may further increase the number of observed, strongly scintillating FRBs, allowing for the stronger constraint of their extragalactic counterparts and Galactic electron distribution models.

\section*{Acknowledgements}
We would like to acknowledge A/Prof. Kiyoshi Masui for refereeing this manuscript and providing helpful insights related to the evolution of spectral modulation indices in bursts undergoing multi-path scattering.
CMT is supported by an Australian Research Council Future Fellowship under project grant FT180100321. ATD acknowledges support from the Australian Government through the Australian Research Council’s Discovery Projects funding scheme (project ID DP220102305). KG acknowledges support through Australian Research Council Discovery Project DP200102243. CWJ and MG acknowledge support by the Australian Government through the Australian Research Council's Discovery Projects funding scheme (DP210102103). RMS acknowledges support through Australian Research Council Future Fellowship FT190100155 and Discovery Project DP220102305.

\section*{Data Availability}

The data underlying this article will be shared upon reasonable request to the corresponding author.



\bibliographystyle{mnras}
\bibliography{references}



\appendix
\begin{landscape}
\begin{table}
\begin{tabular}{|l|l|l|l|l|l|l|l|l|l|l|l|l|}
\hline
FRB & $S/N$ & $m$ & $\nu_c$ (MHz) & $\nu_{DC}$ (MHz) & $t_\text{scatt}$ (ms)& $C$ & $\alpha_\nu$ & $\alpha_t$ & $\nu_\text{Ne2001}$ (MHz) & $\nu_\text{min}$ (MHz) & $L_xL_g$ (kpc$^2$) \\ \hline
   20181112A & 143 & 0.1 & 1297.5 & - & $.0278\pm0.0008$ & -& - & $-0.8\pm0.6$ & 2.82&  0.001 & - &\\ \hline
   20190102C & 124 & 0.41 & 1271.5 & $0.6\pm0.3$ & $0.046\pm0.001$ & 170 & 10 & -4 & 1.26 & 0.001 & - & \\ \hline
   20190608B & 32.9 & 0.78 & 1271.5 &$1.4\pm0.1$  & $4.0\pm0.4$& 35000 & $5.8\pm0.5$ & $-3\pm1$ & 3.08 & 1& $6\pm1$ & \\ \hline
   20190711A & 89.3 & 0.64 & 1136.9 & $0.11\pm0.01$ & $0.008\pm0.003$ & 6  & $-10\pm5$ & -10 & 0.837 & 0.0001 &  & \\ \hline
   20191228A & 51.1 & 0.77 & 1340.3 & - & $5.5\pm0.2$ & - & -& - & 5.47 & 0.1 & - &  \\ \hline
   20200430A & 56.7 & 0.45 & 864.5 & - & $7.7\pm0.5$ & -& - & $-3.0\pm0.3$ &  0.973  & 0.1 & - &  \\ \hline
   20201124A & 172 & 0.59 & 713.9 & $.136\pm0.005$ & $4.04\pm0.07$ & 3450 & $10\pm3$& $-7.3\pm0.9$ & 0.00721  & $1\times10^{-4}$ &  $1.43\pm0.08$&   \\ \hline
   20210117A & 43.9 & 0.0 & 1364.2 & - & $0.14\pm0.6$ & - & - & $1\pm4$ &5.82 & 0.1 & - &  \\ \hline
   20210320C & 113 & 0.83 & 824.2 &  $0.91\pm0.03$ & $0.247\pm0.004$ & 1410& $2\pm1$ & $-3.30\pm0.01$ & 1.03  & 0.1 &  $550\pm30$  &  \\ \hline
   20210407E & 49.7 & 0.0 & 1220.2 & - & $0.090\pm0.010$ & - & - & & $3.01$ & 1 & - &  \\ \hline
\end{tabular}
\caption{Measured scintillation parameters for a sample of localised CRAFT FRBs}
\label{table:burstProperties}
\end{table}
\end{landscape}

\section{Simulating spectral scintillation detection threshold}\label{app:nuMin}
We create a blank spectrum with $336$\,MHz of bandwidth around a central frequency of $1271.5$\,MHz at $0.1$\,MHz resolution. We then populate it with $N=f\times336/\nu_{\text{DC,0}}$ scintles, where the filling fraction $f$ is 0.5, a typical assumption for pulsar scintillation\citep{bhat_long-term_1999, nicastro_scintillation_2001} and $\nu_{\text{DC},0}$ is the decorrelation bandwidth of the simulated burst at $1\,$GHz. The amplitude of the scintles is set such that the sum of the noiseless spectrum is equal to the simulated burst's $S/N$ and their positions in frequency ($\nu_p$) are drawn randomly from a uniform distribution. The scintles are Lorentzian in shape with a HWHM of $2\nu_{\text{DC}}$\citep{bartel_electron_2022}, corresponding to a decorrelation bandwidth of $\nu_{\text{DC}}=\nu_{\text{DC},0}(\nu_p/1\text{\,GHz})^4$. These arrays represent the noise-free signal ($S$) of a burst. We also construct noise arrays ($N$) filled with white noise following a $N(0,1/\sqrt{3360})$ distribution.

\begin{figure}
    \centering
    \includegraphics[width=0.5\textwidth]{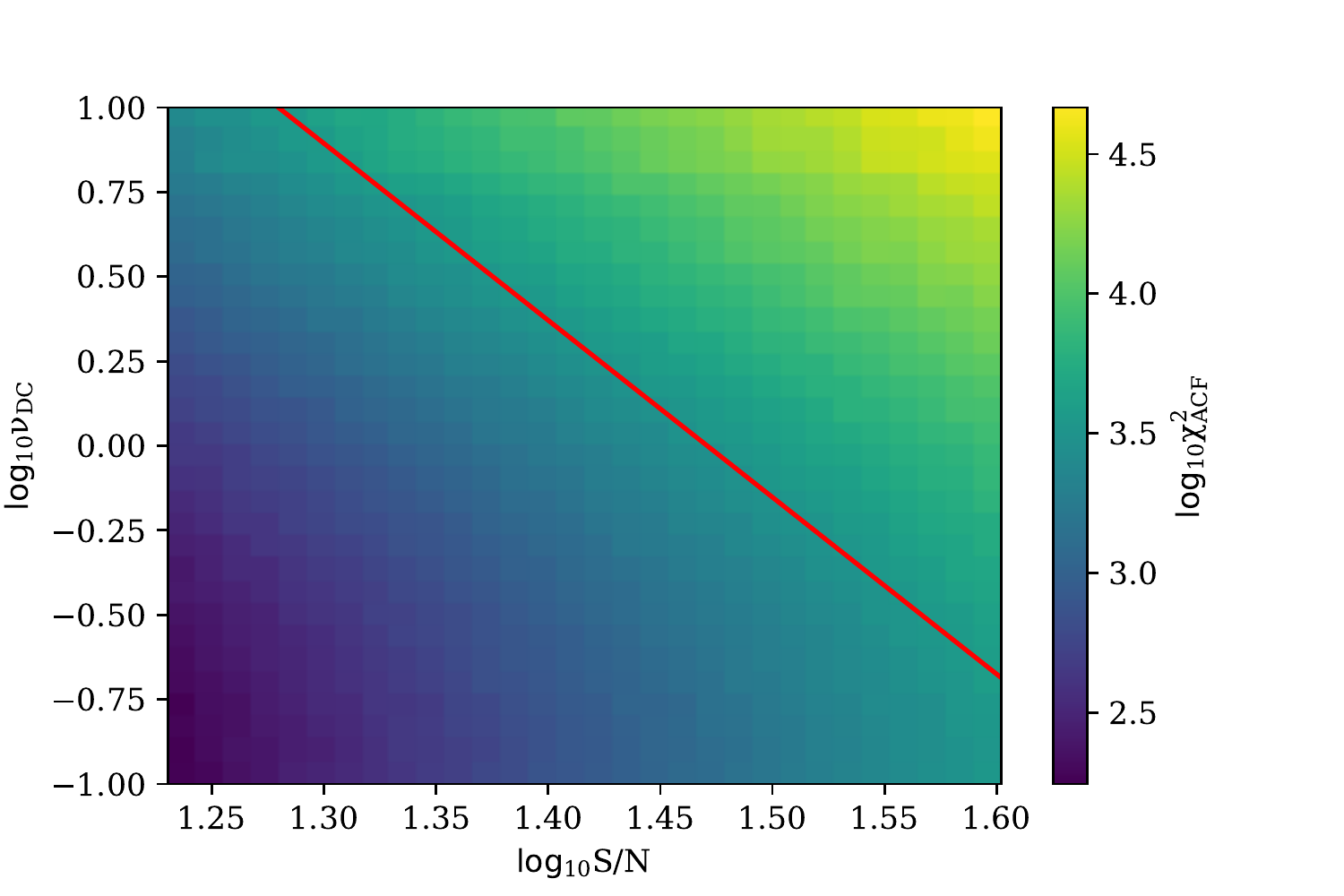}
    \caption{$\log_{10}\chi^2$ values for the significance of scintillation structure in the burst ACF as a function of burst $S/N$ and $\nu_{\text{DC}}$. Red line represents the detection threshold, fit to $\chi^2$ values of $\approx3360$, where the CDF of the $\chi^2$ distribution is $\approx0.5$.}
    \label{fig:nuDcThreshold}
\end{figure}

We simulate 1000 signal and noise arrays for each of a range of combinations of $S/N$ and $\nu_{\text{DC}}$ values. For each, we calculate the ACF signal and noise as
\begin{align}
    \text{ACF}_S(j) &= \sum\limits_{i=0}^{3360}S(i)S(i+j)\\
    \text{ACF}_N(j) &= \sum\limits_{i=0}^{3360}S(i)N(i+j)+S(i+j)N(i)+N(i)N(i+j).
\end{align}
The $\chi^2$ value for the significance of the burst ACF at a given $S/N$ and $\nu_{\text{DC}}$ is then calculated as 
\begin{equation}
    \chi^2 = \sum\limits_{j=0}^{3360}\frac{\overline{\text{ACF}_S(j)}-\overline{\text{ACF}_N(j)}}{\sigma(j)},
\end{equation}
where the bar represents the mean over the 1000 simulated instances and $\sigma(j)$ is the standard deviation of $\text{ACF}_N(j)$. Due to the high degree of freedom of the $\chi^2$ distribution (roughly equal to the number of channels, 3360) the probability of chance significance ($p=1-\text{CDF}(\chi^2)$) also transitions sharply from $\approx1$ to $\approx0$ and hence we set the detection threshold at this transition at $\chi^2$ value of $\approx3360$. Fig. \ref{fig:nuDcThreshold} depicts the simulated $\chi^2$ values and over-plots the detection threshold from linear fit in log space to the $\chi^2\approx3360$ values in red. For values of $\nu_{\text{DC}}$ greater than the threshold at a given $S/N$ the scintillation should be detectable. By extrapolating the threshold relationship we calculate the minimum detectable scintillation bandwidths $\nu_\text{min}$ using the observed burst $S/N$. Given the assumptions used in this model we round $\nu_{\text{min}}$ to the nearest order of magnitude.

\section{Derivation of Eq. 3}\label{app:derivation}
As per \cite{narayan_physics_1992} the modulation index of an extended source is given by 
\begin{align}
    m &= \frac{\theta_{\text{diff}}}{\theta_S}
    \intertext{where $\theta_{\text{diff}}$ is the angle subtended by the diffractive scale $r_{\text{diff}}/D_d$, and $\theta_S$ is the apparent angle of the source. In the case where light scattered by an extragalactic screen into an angle $\theta_{\text{scatt},x}$ is incident upon a Galactic screen characterised by $r_{\text{diff},g}$, the modulation index of scintillation from the Galactic screen ($m_g$) will be given by}
    m_g &= \frac{r_{\text{diff},g}}{D_{d,g}\theta_{\text{scatt},x}}
    \intertext{where $D_{d,g}$ is the distance to the Galactic screen from the observer. The diffractive scale the Galactic screen may also be approximated as $r_{\text{diff},g}\sim\lambda /2\pi \theta_{\text{scatt},g}$ \citep{narayan_physics_1992}, yielding}
    m_g &= \frac{\lambda}{2\pi\theta_{\text{scatt},g}\theta_{\text{scatt},x}D_{d,g}}
    \intertext{where $\lambda$ is the observed wavelength. As per \cite{macquart_temporal_2013} scattering angles can be expressed as scattering times following $t_{\text{scatt}}=D_dD_s\theta_\text{scatt}^2/[cD_{ds}(1+z_d)]$, where $z_d$ is the redshift at the screen. Substituting the scattering angles for scattering times gives}
    t_{\text{scatt},g}t_{\text{scatt},x} &= \frac{1}{(2\pi\nu)^2m_g^2(1+z_x)}\frac{D_{s,x}D_{d,x}}{D_{ds,x}}\frac{D_{s,g}}{D_{ds,g}D_{d,g}}\label{eq:generalTwoScreen}
    \intertext{When $m_g=1$, the left hand side becomes $\lesssim$ the right hand side, and the general form of Eq. \ref{eq:twoScreenDist} is recovered. By assuming that the screens are close to their respective ends of the path length (i.e. they are associated with the Milky Way and host galaxies), we can approximate $D_{d,x}\approx D_{ds,g}$, and $D_{s,x}\approx D_{s,g}\approx D_s$, reducing the above expression to}
    t_{\text{scatt},g}t_{\text{scatt},x}&\approx \frac{D_{s}^2}{(2\pi\nu)^2m_g^2}\frac{1}{D_{d,g}D_{ds,x}}
    \intertext{Exchanging $t_{\text{scatt,g}}$ for $1/2\pi\nu_{\text{DC}}$ via Eq. \ref{eq:uncertainty} and recognising that $D_{d,g}$ and $D_{ds,x}$ are $L_g$ and $L_x$ respectively yields Eq. \ref{eq:PartiallyResolved2ScreenScat}}
    L_xL_g &\approx \frac{D_s^2}{2\pi\nu^2(1+z_x)}\frac{\nu_{\text{DC}}}{t_{\text{scatt}}}
\end{align}

\section{Anisotropic Scattering Screens}\label{sec:anisotropy}
The constraints on the $L_xL_g$ product are derived from the condition, that to have fully modulated diffractive scintillation at the second screen, the coherence length set by the diffractive scale of the first scattering screen ($r_{\text{diff},1}$) must be greater than the scattering angle of the second screen projected back onto that screen \citep{ocker_large_2022}, i.e.
\begin{equation}\label{eq:coherenceLengthCond}
    r_{\text{diff,1}}\geq\theta_\text{scatt,2}D_d.
\end{equation}

The implicit assumption within this condition is that the thin screens are two-dimensional and isotropic. Under this assumption the extent of angular broadening caused by the first screen is equivalent to the extent of the source as seen by the second screen. If however, the screens were anisotropic, the direction of angular broadening will also be important. In the extreme case where each screen is one dimensional, e.g. scattering by a filament or tidal stream similar to that observed by \cite{wang_askap_2021}, then the angular extent of the source seen by the second screen will be given by the projection of the image scattered by the first screen onto the second. The condition to observe fully modulated diffractive scintillation at the second screen then becomes 
\begin{equation}
    r_{\text{diff},1}\geq\theta_\text{scatt,2}D_d\cos{\phi},
\end{equation}
where $\phi$ would be the angle between the one-dimensional screens if they were projected onto a plane perpendicular to the optic axis. 

For parallel, one-dimensional screens the constraints will remain unchanged. Conversely, for perfectly orthogonal screens, the second screen will always observe the source to be point-like, regardless of the extent of angular broadening from the first screen. In this case, the scattering in each dimension will be completely independent and we will be unable to constrain the scattering geometry. Constraints on the two-screen distance product are therefore completely degenerate with the anisotropy of the scattering screens. 

We highlight, however, that in the case of anisotropic scattering screens, it is expected that the shapes of both the temporal impulse response function (i.e. the temporal broadening profile) and the spectral ACF of the scattered pulse will differ from those used here. Specifically, in the case of the temporal impulse response function, anisotropic screens are expected to show a greater fraction of intensity at larger time delays \citep{cordes_anomalous_2001, rickett_anisotropy_2006}. The more one-dimensional these screens become, the greater the difference in expected pulse morphology will be. Within our sample, the observed pulses are well described by the pulse morphology expected for scattering through an isotropic, two-dimensional thin screen, particularly FRB\,20210320C. We, therefore, conclude that significant anisotropy in the scattering screens associated with our observations is unlikely, and we leave a rigorous treatment of the expected FRB morphology for scattering through anisotropic screens to a future work.

\section{Additional Figures}
\begin{figure}
    \centering
\includegraphics[width=0.5\textwidth]{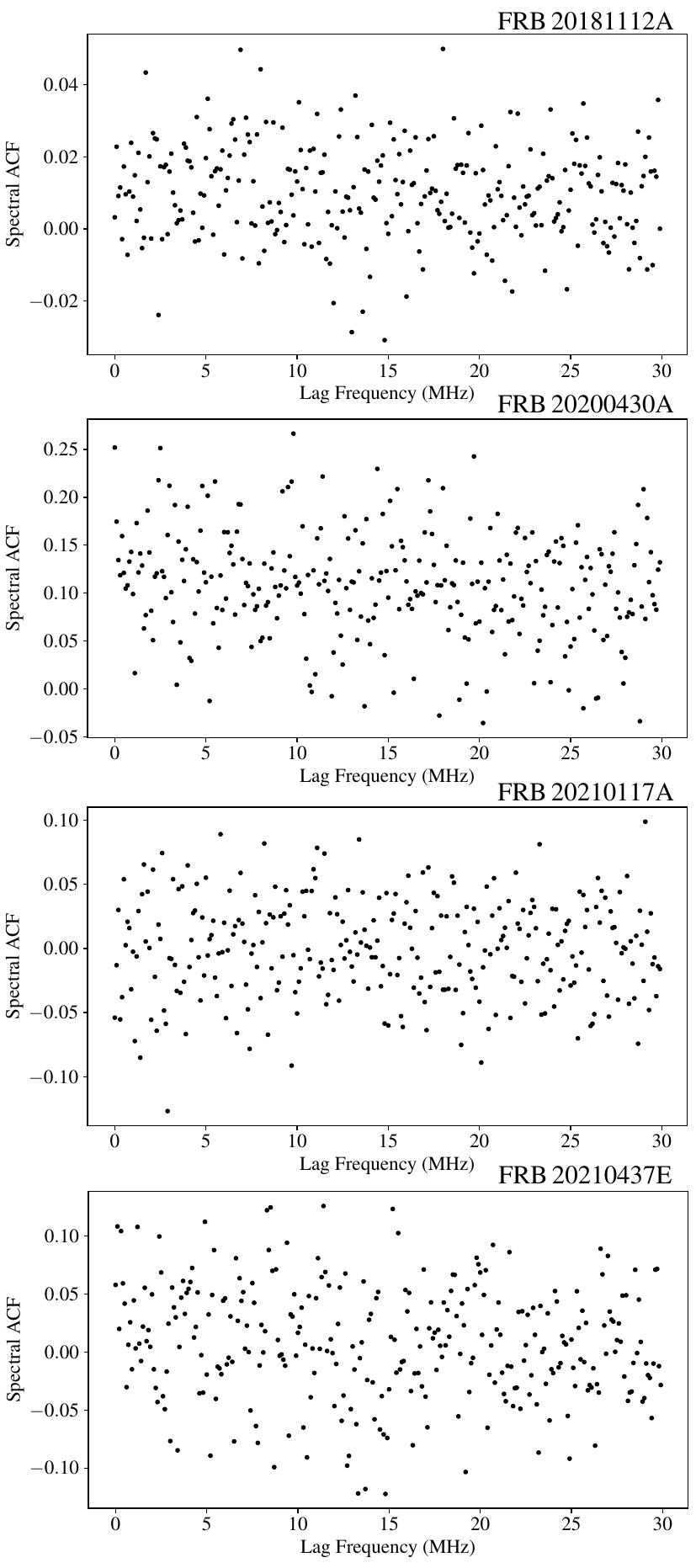}
    \caption{Time-integrated spectral ACFs for FRBs within the sample confirmed not to scintillate. Each panel is labelled with the relevant FRB name.}
    \label{fig:NoScint}
\end{figure}

\begin{figure*}
    \centering
    \includegraphics[width=\textwidth]{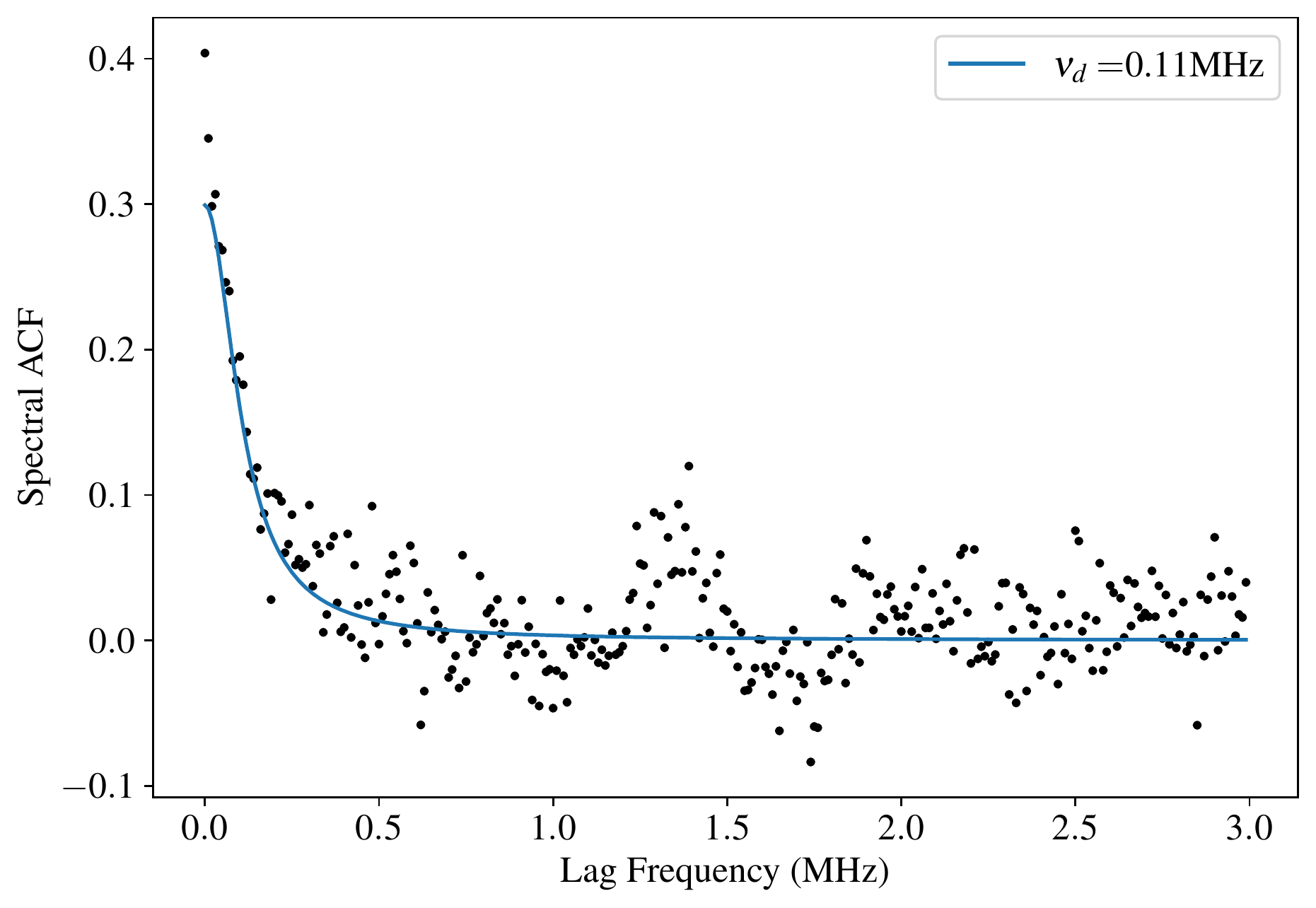}
    \caption{Time-integrated spectral ACF for FRB\,20190711A, for which we found insufficient evidence to prove scintillation.}
    \label{fig:Strange190711}
\end{figure*}

\begin{figure*}
    \centering
    \includegraphics[width=\textwidth]{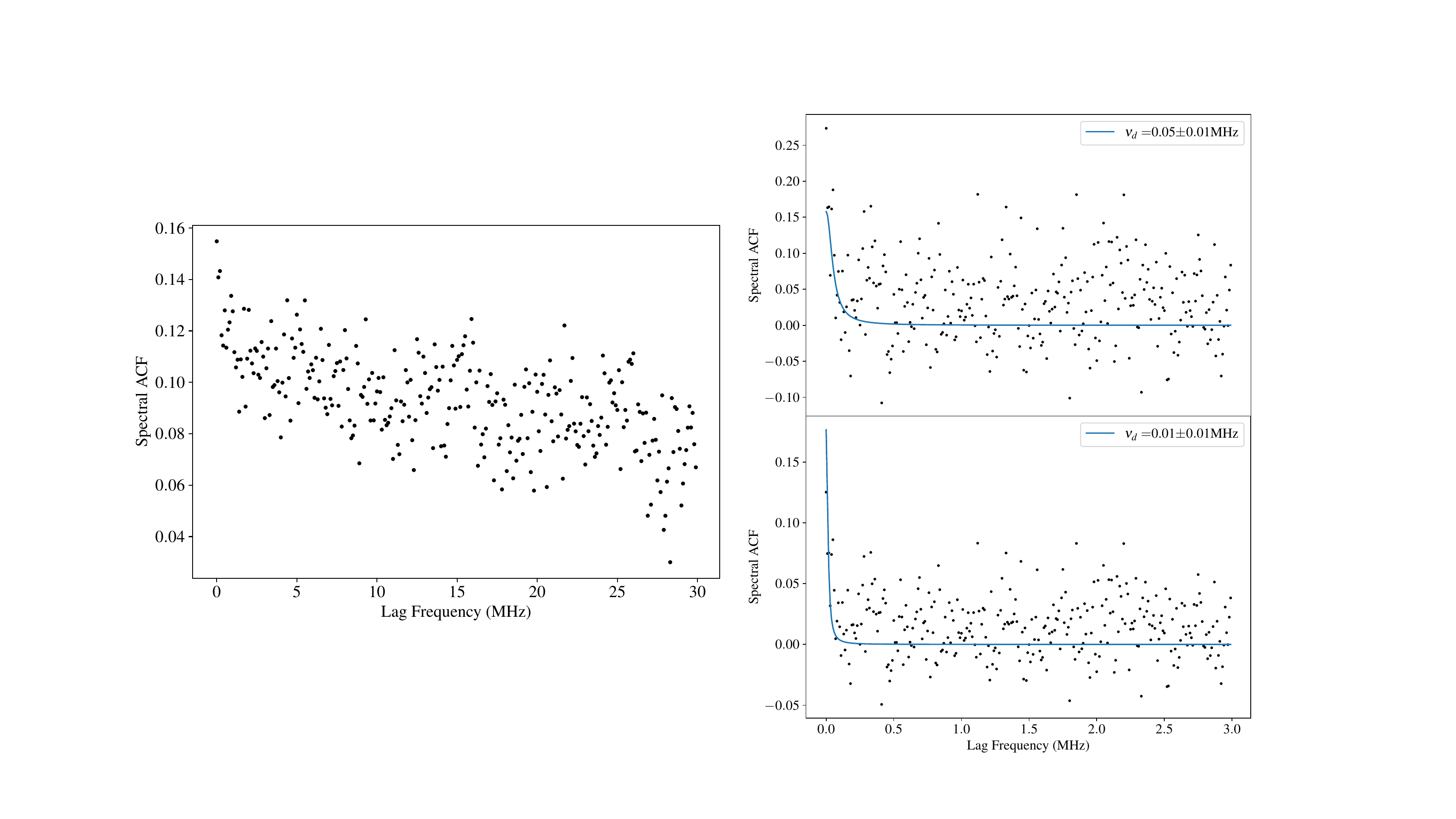}
    \caption{Time-integrated spectral ACFs for FRB\,20190102C, for which we found insufficient evidence to prove scintillation. \textit{Left}: ACF over the full used bandwidth. \textit{Right}: ACFs for the \textit{top} and \textit{bottom} sub-bands respectively.}
    \label{fig:Strange190102}
\end{figure*}

\begin{figure*}
    \centering
    \includegraphics[width=\textwidth]{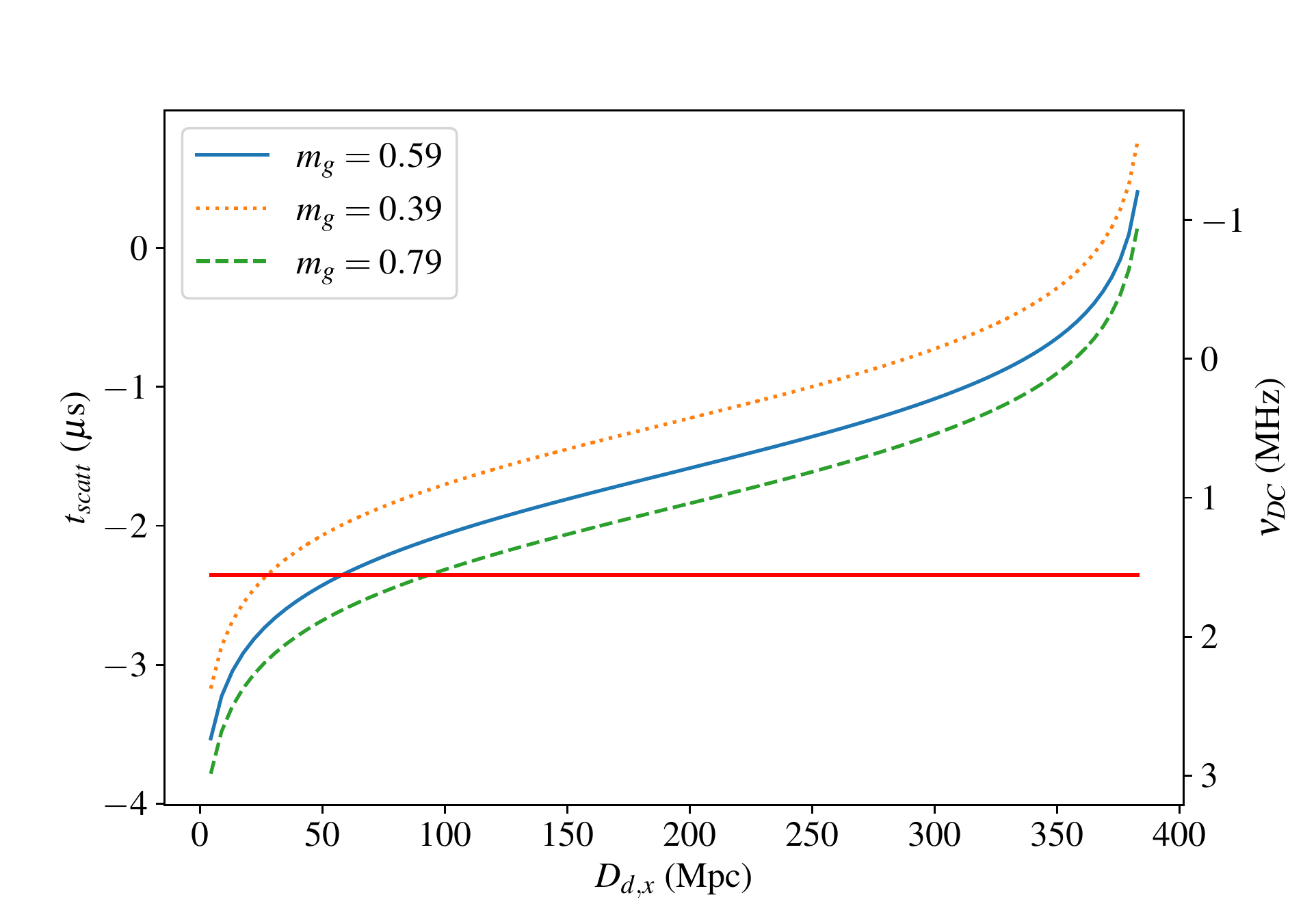}
    \caption{Scattering timescale and corresponding decorrelation bandwidths (assuming $2\pi\nu_{DC}t_{\text{scatt}}=1$) contributed by a potential third IGM screen which would be responsible for suppressing the modulation of spectral scintillation in FRB\,20201124A through angular broadening. The blue line corresponds to the calculation for the measured modulation index of $m_g=0.59$, additional dotted and dashed lines display how the required scattering time changes with the modulation index. The red line displays the observed bandwidth of FRB\,20201124A.}
    \label{fig:20201124AIGM}
\end{figure*}

\begin{figure*}
    \centering
    \includegraphics[width=\textwidth]{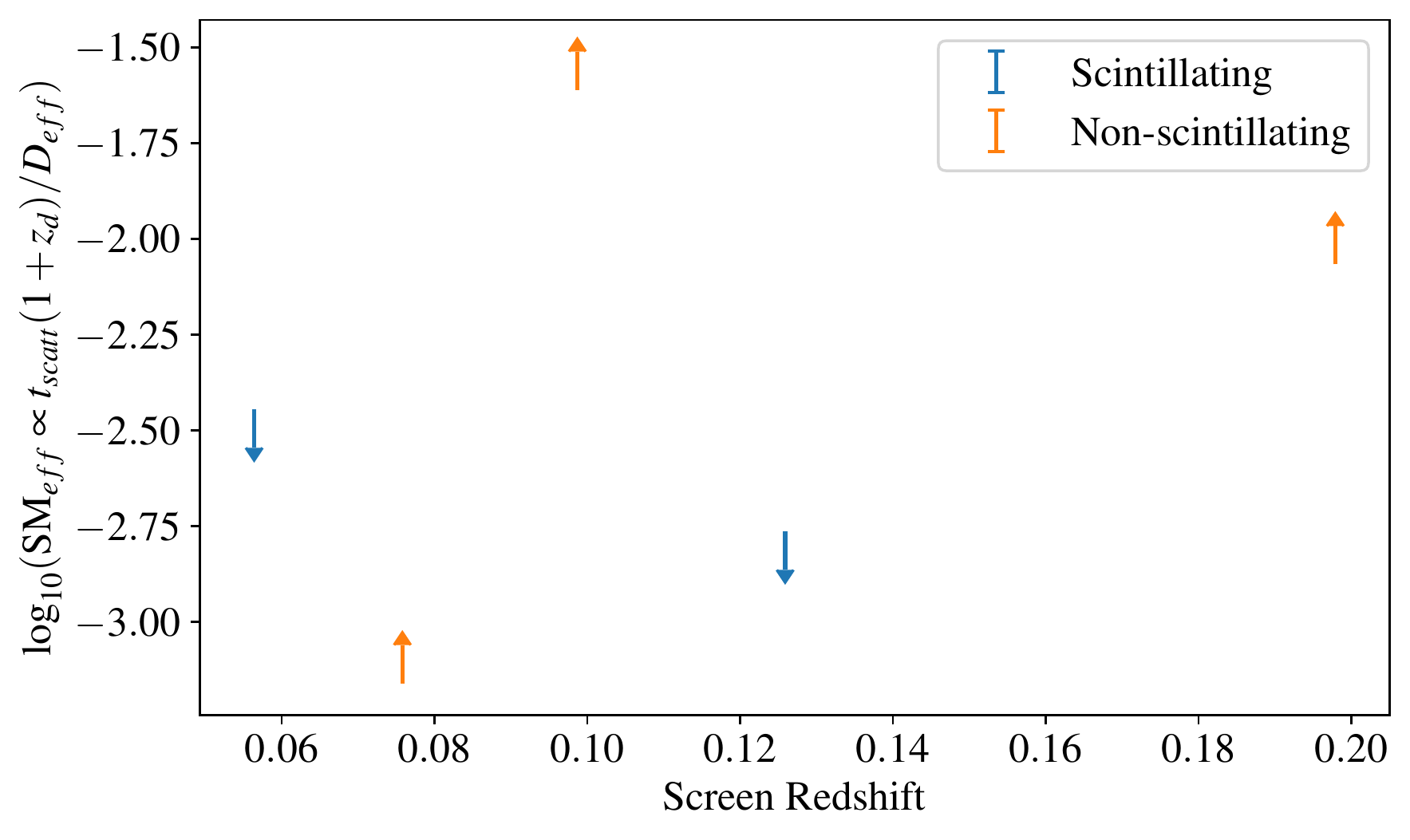}
    \caption{Limits on values proportional to the effective scattering measure defined in \citet{macquart_temporal_2013} from scintillating and non-scintillating FRBs as discussed in \S \ref{subsec:190608Dis} and \ref{subsec:210320Dis} and \S \ref{subsec:galacticScint} respectively.}
    \label{fig:AllSMConst}
\end{figure*}

\bsp	
\label{lastpage}
\end{document}